\documentclass[prx,twocolumn,floatfix,superscriptaddress,longbibliography]{revtex4-1}
\usepackage[utf8]{inputenc}
\usepackage[T1]{fontenc}
\usepackage{bm}
\usepackage{tikz}
\usepackage{subfiles}
\usepackage{mathtools}
\usetikzlibrary{shapes,arrows,positioning}
\usepackage{standalone}
\usepackage{amssymb,amsmath,amstext}
\usepackage{amsthm}
\usepackage{bbm}
\usepackage{graphicx}
\usepackage{epstopdf}
\usepackage{color}
\usepackage{appendix}
\usepackage[T1]{fontenc}
\usepackage{bbold}
\usepackage{bbm}
\usepackage{float}
\usepackage{latexsym}
\usepackage{xr-hyper}
\usepackage[colorlinks=true,citecolor=blue,linkcolor=magenta]{hyperref}

\usepackage{hhline}
\usetikzlibrary{shapes,arrows}
\renewcommand{\vec}[1]{\boldsymbol{#1}}  % Bold vectors instead of arrow vectors

\long\def\ca#1\cb{} %Use for commenting out: \ca...\cb

\newcommand{\ket}[1]{|#1\rangle}               %ket
\newcommand{\bra}[1]{\langle #1|}              %bra
\newcommand{\dya}[1]{\ket{#1}\!\bra{#1}}
\newcommand{\OC}{\mathcal{O}}

\newcommand{\Tr}{{\rm Tr}}

\renewcommand{\geq}{\geqslant}

\DeclareMathOperator*{\argmin}{arg\,min}
\renewcommand{\vec}[1]{\boldsymbol{#1}}  % Bold vectors instead of arrow vectors

\newcommand{\ad}{^\dagger}
\newcommand*{\id}{\mathbbm{1}}

\begin{document}

\title{Variational quantum algorithms}

\author{M. Cerezo}
\thanks{e-mail: cerezo@lanl.gov}
\affiliation{Theoretical Division, Los Alamos National Laboratory, Los Alamos, NM 87545, USA}
\affiliation{Center for Nonlinear Studies, Los Alamos National Laboratory, Los Alamos, NM, USA}
\affiliation{Quantum Science Center, Oak Ridge, TN 37931, USA}

\author{Andrew Arrasmith}
\affiliation{Theoretical Division, Los Alamos National Laboratory, Los Alamos, NM 87545, USA}
\affiliation{Quantum Science Center, Oak Ridge, TN 37931, USA}

\author{Ryan Babbush}
\affiliation{Google Quantum AI Team, Venice, CA 90291, United States of America}

\author{Simon C. Benjamin}
\affiliation{Department of Materials, University of Oxford, Parks Road, Oxford OX1 3PH, United Kingdom}

\author{Suguru Endo}
\affiliation{NTT Secure Platform Laboratories, NTT Corporation, Musashino, Tokyo 180-8585, Japan}

\author{Keisuke Fujii}
\affiliation{Graduate School of Engineering Science, Osaka University, Osaka 560-8531, Japan}
\affiliation{Center for Quantum Information and Quantum Biology, Institute for Open and Transdisciplinary Research Initiatives, Osaka University, Osaka 560-8531, Japan}
\affiliation{Center for Emergent Matter Science, RIKEN, Saitama 351-0198, Japan}

\author{Jarrod R. McClean}
\affiliation{Google Quantum AI Team, Venice, CA 90291, United States of America}

\author{Kosuke Mitarai}
\affiliation{Graduate School of Engineering Science, Osaka University, Osaka 560-8531, Japan}
\affiliation{Center for Quantum Information and Quantum Biology, Institute for Open and Transdisciplinary Research Initiatives, Osaka 560-8531, Japan}
\affiliation{JST, PRESTO, Saitama 332-0012, Japan}

\author{Xiao Yuan}
\affiliation{Center on Frontiers of Computing Studies, Department of Computer Science, Peking University, Beijing 100871, China}
\affiliation{Stanford Institute for Theoretical Physics, Stanford University, Stanford California 94305, USA}

\author{Lukasz Cincio}
\affiliation{Theoretical Division, Los Alamos National Laboratory, Los Alamos, NM 87545, USA}
\affiliation{Quantum Science Center, Oak Ridge, TN 37931, USA}

\author{Patrick J. Coles}
\thanks{e-mail: pcoles@lanl.gov}
\affiliation{Theoretical Division, Los Alamos National Laboratory, Los Alamos, NM 87545, USA}
\affiliation{Quantum Science Center, Oak Ridge, TN 37931, USA}

\begin{abstract}
Applications such as simulating complicated quantum systems or solving large-scale linear algebra problems are very challenging for classical computers due to the extremely high computational cost. Quantum computers promise a solution, although fault-tolerant quantum computers will likely not be available in the near future. Current quantum devices have serious constraints, including limited numbers of qubits  and noise processes that limit circuit depth. Variational Quantum Algorithms (VQAs), which use a classical optimizer to train a parametrized quantum circuit, have emerged as a leading strategy to address these constraints. VQAs have now been proposed for essentially all applications that researchers have envisioned for quantum computers, and they appear to the best hope for obtaining quantum advantage. Nevertheless, challenges remain including the trainability, accuracy, and efficiency of VQAs. Here we overview the field of VQAs, discuss strategies to overcome their challenges, and highlight the exciting prospects for using them to obtain quantum advantage.
\end{abstract}

\maketitle

\section{Introduction}
Quantum computing holds promise for a number of applications that have motivated the decades-long quest to build the necessary physical hardware. For example, with an exponential speedup over classical methods, quantum algorithms could factor numbers~\cite{shor1994algorithms}, simulate quantum systems ~\cite{lloyd1996universal}, or solve linear systems of equations~\cite{harrow2009quantum}. 

In 2016, access to the first cloud-based quantum computer~\cite{IBM} became available, but noise and qubit limitations prevented serious implementations of the aforementioned quantum algorithms~\cite{coles2018quantum}. However, excitement grew as to what could be done with these new devices, which have been called Noisy Intermediate-Scale Quantum (NISQ) computers~\cite{preskill2018quantum}. Current state-of-the-art device size ranges from 50 to 100 qubits which allows one to achieve `quantum supremacy': outperforming the best classical supercomputer, for certain contrived mathematical tasks~\cite{arute2019quantum,zhong2020quantum}. 

Nevertheless, the true promise of quantum computers, speedup for practical applications, which is often called quantum advantage, has yet to be realized. Moreover, the availability of fault-tolerant quantum computers appears to still be many years, or even decades, away. The key technological question is therefore how to make best use of today's NISQ devices to achieve quantum advantage. Any such strategy must account for: limited numbers of qubits, limited connectivity of the qubits, and coherent and incoherent errors that limit quantum circuit depth. 

\begin{figure*}[t!]
    \centering
    \includegraphics[width=1\linewidth]{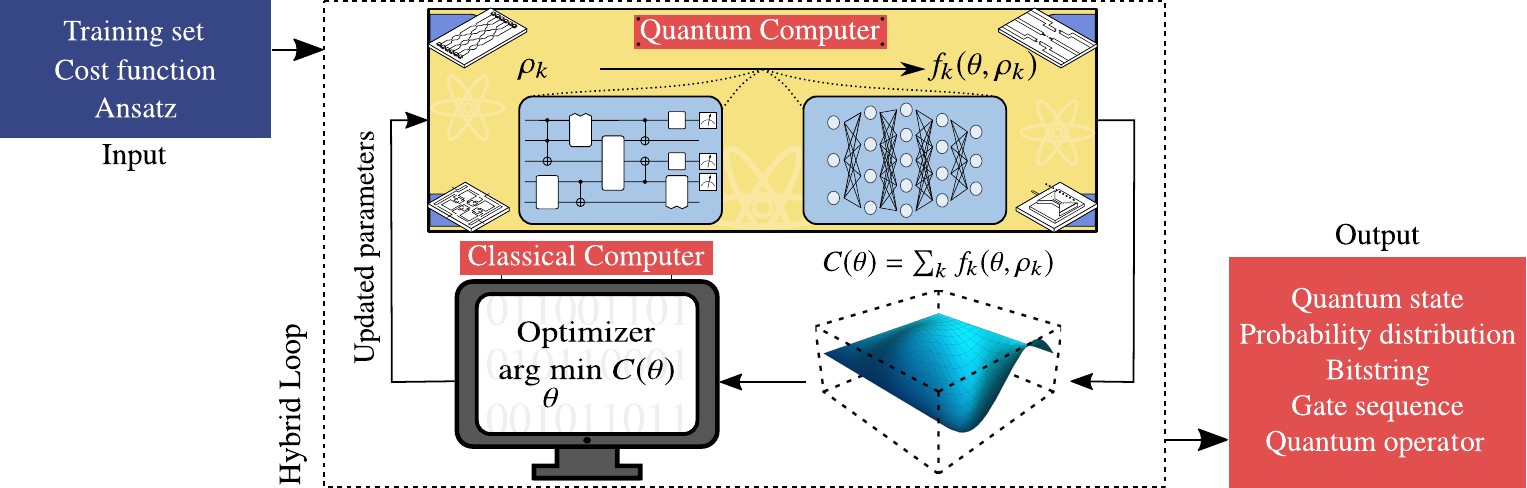}
    \caption{\textbf{Schematic diagram of a Variational Quantum Algorithm (VQA)}. The inputs to a VQA are: a cost function $C(\vec{\theta})$, with $\vec{\theta}$ a set of parameters that encodes the solution to the problem, an ansatz whose parameters are trained to minimize the cost, and (possibly) a set of training data $\{\rho_k\}$ used during the optimization. Here, the cost can often be expressed in the form in Eq.~\eqref{eq:cost-simple}, for some set of functions $\{f_k\}$. Also, the ansatz is shown as a parameterized quantum circuit (on the left), which is analogous to a neural network (also shown schematically on the right). At each iteration of the loop one uses a quantum computer to efficiently estimate the cost (or its gradients). This information is fed into a classical computer that leverages the power of optimizers to navigate the cost landscape $C(\vec{\theta})$ and solve the optimization problem in Eq.~\eqref{eq:optimization}. Once a termination condition is met, the VQA outputs an estimate of the solution to the problem. The form of the output depends on the precise task at hand. The red box indicates some of the most common types of outputs. }
    \label{fig:VQA}
\end{figure*}

Variational Quantum Algorithms (VQAs) have emerged as the leading strategy to obtain quantum advantage on NISQ devices. Accounting for all of the constraints imposed by NISQ computers with a single strategy requires an optimization-based or learning-based approach, precisely what VQAs use. VQAs are arguably the quantum analog of highly successful machine-learning methods, such as neural networks. Moreover, VQAs leverage the toolbox of classical optimization, since VQAs use parametrized quantum circuits to be run on the quantum computer, and then outsource the parameter optimization to a classical optimizer. This approach has the added advantage of keeping the quantum circuit depth shallow and hence mitigating noise, in contrast to quantum algorithms developed for the fault-tolerant era.

VQAs have already been considered for a plethora of applications (see Figure~\ref{fig:Appli}), covering essentially all of the applications that researchers had envisioned for quantum computers. Although they may be the key to obtaining near-term quantum advantage, VQAs still face important challenges, including their trainability, accuracy, and efficiency. In this Review, we discuss the exciting prospects for VQAs, and we highlight the challenges that must be overcome to obtain the ultimate goal of quantum advantage.

\section{Basic concepts and tools}\label{sec:2}

One of the main advantages of VQAs is that they provide a general framework that can be used to solve a variety of problems. Although this versatility translates into different algorithmic structures with different levels of complexity, there are basic elements that most (if not all) VQAs have in common. In this section we review the building blocks of VQAs.

Let us start by considering a task one wishes to solve. This implies having access to a description of the problem, and also possibly to a set of training data. As schematically shown in Fig.~\ref{fig:VQA}, the first step to developing a VQA is to define a cost (or loss) function $C$ which encodes the solution to the problem. One then proposes an ansatz, that is, a quantum operation depending on a set of continuous or discrete parameters  $\vec{\theta}$ that can be optimized (see below for a more in-depth discussion of ansatzes). This ansatz is then 
trained  in a hybrid quantum-classical loop to solve the optimization task 
\begin{equation}\label{eq:optimization}
   \vec{\theta}^*= \argmin_{\vec{\theta}} C(\vec{\theta})\,.
\end{equation}
The trademark of VQAs is that they use a quantum computer to estimate the cost function $C(\vec{\theta})$ (or its gradient) while leveraging the power of classical optimizers to train the parameters $\vec{\theta}$. In what follows, we provide additional details for each step of the VQA architecture shown in Fig.~\ref{fig:VQA}.

\subsection{ Cost function}\label{sec:Cost}

A crucial aspect of a VQA is encoding the problem into a cost function. Similar to classical machine learning, the cost function maps values of the trainable parameters $\vec{\theta}$ to real numbers. More abstractly, the cost defines a hyper-surface usually called the cost landscape (see Fig.~\ref{fig:VQA}) such that the task of the optimizer is to navigate through the landscape and find the global minima. Without loss of generality, the cost can be expressed as
\begin{equation}\label{eq:cost-general}
    %C(\vec{\theta}) = \sum_{k}f_k\left(\Tr[O_kU(\vec{\theta})\rho_k U\ad(\vec{\theta})]\right)\,.
    C(\vec{\theta}) = f\left(\{\rho_k\},\{O_k\},U(\vec{\theta})\right)\,,
\end{equation}
where $f$ is some function, $U(\vec{\theta})$ is a parametrized unitary,  $\vec{\theta}$ is composed of discrete and continuous parameters, $\{\rho_k\}$ are input states from a training set, $\{O_k\}$ are a set of observables. Often it is useful, and possible, to express the cost in the form
\begin{equation}\label{eq:cost-simple}
    C(\vec{\theta}) = \sum_{k}f_k\left(\Tr[O_kU(\vec{\theta})\rho_k U\ad(\vec{\theta})]\right)\,,
%    C(\vec{\theta}) = f\left(\{\rho_k\},\{O_k\},U(\vec{\theta})\right)\,,
\end{equation}
for some set of functions $\{f_k\}$. Note that the task at hand will determine the choice of $f$ in Eq. \eqref{eq:cost-general} or the choice of $\{f_k\}$ in Eq. \eqref{eq:cost-simple}. During the optimization, one uses a finite statistic estimator of the cost or its gradients. (See below  for an overview of optimizers used to train the cost function.)

Let us now discuss desirable criteria that the cost function should meet. First, the cost must be  `faithful' in that the minimum of  $C(\vec{\theta})$ corresponds to the solution of the problem. Second, one must be able to  `efficiently estimate' $C(\vec{\theta})$ by performing measurements on a quantum computer and possibly performing classical post-processing. An implicit assumption here is that  the cost should not  be efficiently computable with a classical computer, as this would imply that no quantum advantage can be achieved with the VQA. In addition, it is also useful for  $C(\vec{\theta})$ to be `operationally meaningful', so that smaller cost values indicate a better solution quality. Finally, the cost must be `trainable', which means that it should be possible to efficiently optimize the parameters $\vec{\theta}$. We will later discuss in more detail the issue of trainability for VQAs.

For a given VQA to be implementable in NISQ hardware, the quantum circuits used to estimate $C(\vec{\theta})$ must  keep the  circuit depth and ancilla requirements small. This is due to the fact that NISQ devices  are prone to gate errors, have limited qubit counts, and that these qubits have short decoherence times. Hence the construction of efficient cost evaluation circuits is an important aspect of VQA research.

\subsection{Ansatzes}\label{sec:ansatzes}

\begin{figure}[t]
    \centering
    \includegraphics[width=1\columnwidth]{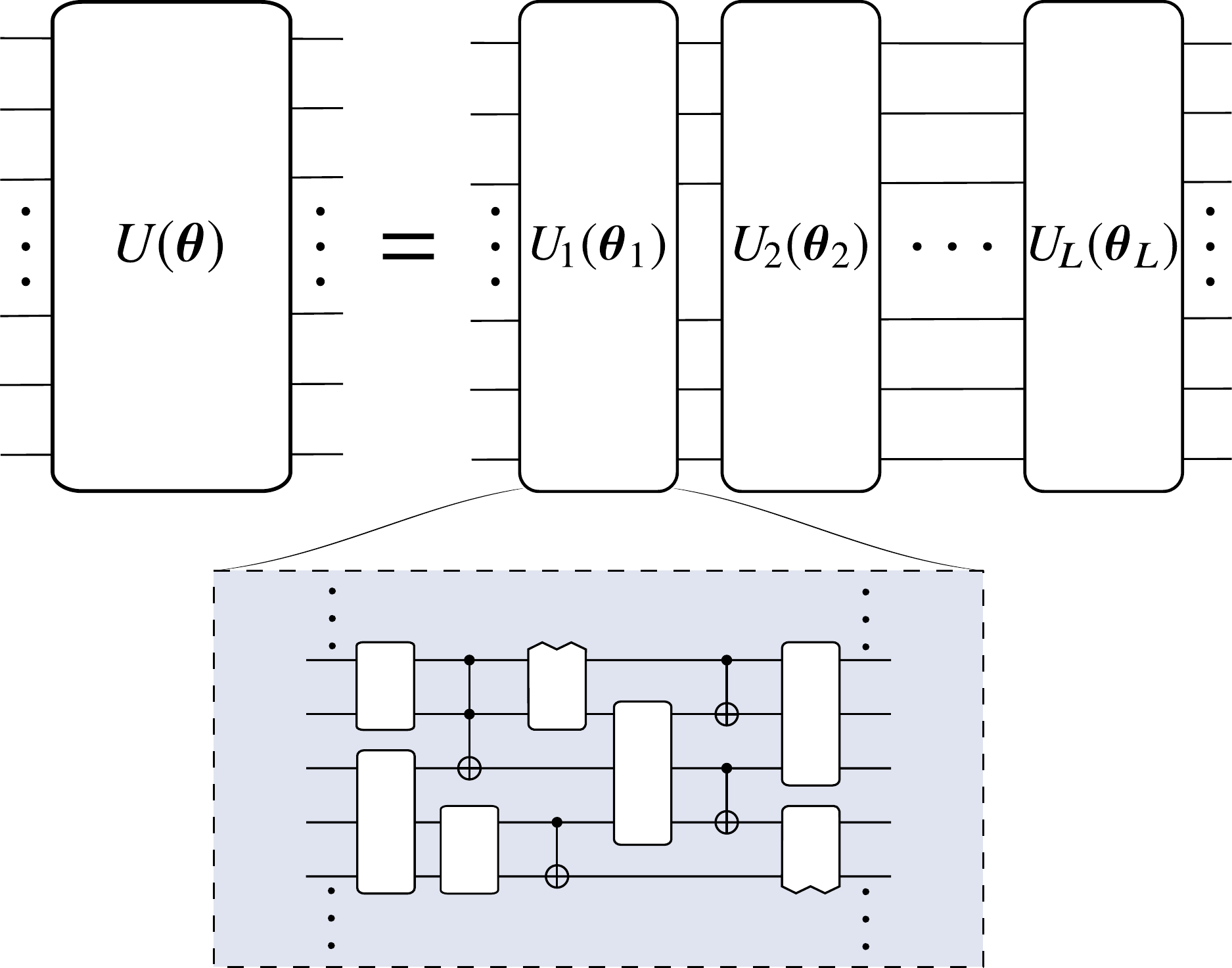}
    \caption{\textbf{Schematic diagram of an ansatz}. The unitary $U(\vec{\theta})$, with $\vec{\theta}$ a set of parameters, can be expressed as a product of $L$ unitaries $U_l(\vec{\theta}_l)$ sequentially acting on an input state. As indicated, each unitary $U_l(\vec{\theta}_l)$ can in turn be decomposed into a sequence of parametrized and  unparametrized gates.  }
    \label{fig:ansatz}
\end{figure}

Another important aspect of a VQA is its ansatz. Generically speaking the form of the ansatz dictates what the parameters $\vec{\theta}$ are, and hence, how they can be trained to minimize the cost. The specific structure of an ansatz will generally depend on the task at hand, as in many cases one can use information about the problem to tailor an ansatz. These are the so-called `problem-inspired ansatze'. However, some ansatz architectures are generic and `problem-agnostic', meaning that they can be used even when no relevant information is readily available. For the cost function in Eq.~\eqref{eq:cost-simple}, the parameters $\vec{\theta}$ can be encoded in a unitary $U(\vec{\theta})$ that is applied to the input states to the quantum circuit. As shown in Fig.~\ref{fig:ansatz}, $U(\vec{\theta})$ can be generically expressed as the product of $L$ sequentially applied unitaries 
\begin{equation}\label{eq:general-ansatz-chain}
U(\vec{\theta})=U_L(\vec{\theta}_L)\cdots U_2(\vec{\theta}_2)U_1(\vec{\theta}_1)\,,
\end{equation}
with
\begin{equation}
U_l(\theta_l)= \prod_{m} e^{-i \theta_{m} H_{m}} W_{m}\,.
\end{equation}
Here $W_{m}$ is an unparametrized unitary  and $H_{m}$ is a Hermitian operator; $\theta_l$ is the $l$-th element in $\vec{\theta}$. Below we describe some of the most widely used ansatzes in the literature, starting with those that can be expressed as Eq.~\eqref{eq:general-ansatz-chain}, and then presenting more general architectures.

\subsubsection{Hardware efficient ansatz}  The hardware efficient ansatz~\cite{kandala2017hardware} is a generic name used for ansatzes that are aimed at reducing the circuit depth needed to implement $U(\vec{\theta})$ when using a given quantum hardware. Here one uses unitaries $W_m$ and $e^{-i \theta_{m} H_{m}}$ that are taken from a gate alphabet (set of quantum gates) determined from the connectivity and interactions specific to a quantum hardware which avoids the circuit depth overhead arising from  translating an arbitrary unitary into a sequence of gates easily implementable in a device.   One of the main advantages of the hardware efficient ansatz is its versatility, as it can accommodate encoding symmetries~\cite{gard2020efficient,otten2019noise} and bringing correlated qubits closer for depth reduction~\cite{tkachenko2020correlation}, as well as being especially useful to study Hamiltonians that are similar to the device's interactions~\cite{kokail2019self}. Such is the case, for instance, of local spin Hamiltonians, although in this case it has been heuristically shown that  near criticiallity the ansatz requires depths proportional to the system size~\cite{bravo2020scaling}. Additionally, `layered' hardware efficient ansatzes, where gates act on alternating pairs of qubits in a brick-like structure, have been prominently used as problem-agnostic architectures.  However, this ansatz can lead to trainability problems when randomly initialized.

\subsubsection{Unitary coupled clustered ansatz}  The Unitary Coupled (UCC) ansatz is a problem-inspired ansatz widely used in quantum chemistry problems where the goal is to obtain the ground state energy of a fermionic molecular Hamiltonian $H$.  The UCC ansatz proposes a candidate for such ground state based on  exciting some reference state $\ket{\psi_0}$ (usually the Hartree-Fock state of $H$) as $e^{T(\vec{\theta})-T(\vec{\theta})\ad}\ket{\psi_0}$. Here, $T=\sum_k T_k$ is the cluster operator~\cite{Taube2006ucc,VQE} and $T_k$ are excitation operators. In the so-called UCCSD ansatz (SD stands for single and double) the summation is truncated  to contain single excitations $T_{1}= \sum_{i,j} \theta_{i}^{j} a_{i}^{\dagger} a_{j}$, and double excitations $T_{2}= \sum_{i, j,k,l} \theta_{i, j}^{k,l} a_{i}^{\dagger} a_{j}^{\dagger} a_{k} a_{l}$, where $\{a_i\ad\}$  ($\{a_i\}$)  are fermionic creation (annihilation) operators.  
To implement this ansatz  in a quantum computer one uses the Jordan-Wigner or the Bravyi-Kitaev transformations~\cite{bravyi2002fermionic} to map the fermionic operators to spin operators, resulting in an ansatz of the form Eq.~\eqref{eq:general-ansatz-chain}.
There are many variants of the UCC ansatz~\cite{Lee2019generalized}, with some of them reducing the circuit depth by considering more efficient methods for compiling the fermionic operators~\cite{motta2018low,Matsuzawa2020,kivlichan2018quantum,setia2019superfast}.

\subsubsection{Quantum alternating operator ansatz} The Quantum Approximate Optimization Algorithm (QAOA) was originally introduced to obtain approximate solutions for combinatorial optimization problems~\cite{qaoa2014}. The ansatz used in QAOA involves an alternating structure and is often called the quantum alternating operator ansatz~\cite{nasaQAOA2019}, sharing the same acronym as the algorithm (although we will use QAOA to refer to the algorithm in this Review). This ansatz was first shown to be computationally universal for certain Hamiltonians in Ref.~\cite{Lloyd_2018}, with the proof of its universality being generalized in Ref.~\cite{morales2020universality} for families  of ansatzes defined by sets of graphs and hyper-graphs.   The ansatz in QAOA is inspired by a Trotterized adiabatic transformation where the order $p$ of the Trotterization determines the precision of the solution. The goal of this ansatz is to map an input state $\ket{\psi_0}$ to the ground-state of a given problem Hamiltonian $H_P$ by  sequentially applying a problem unitary $e^{-i \gamma_l H_p}$ and a mixer unitary $e^{-i \beta_l H_M}$, where $H_M$ is a Hermitian operator known as the mixing Hamiltonian Ref.~\cite{wang2020x}. Specifically, the ansatz takes the form $U(\vec{\gamma},\vec{\beta})=\prod_{l=1}^p e^{-i \beta_l H_M}e^{-i \gamma_l H_P}$, where $\vec{\theta}=(\vec{\gamma},\vec{\beta})$. This ansatz is naturally of the form in Eq.~\eqref{eq:general-ansatz-chain}, although decomposing these unitaries into native gates may result in a lengthy circuit due to many-body terms in $H_P$ and limited device connectivity. One of the strengths of this ansatz is the fact that the feasible subspace for certain problems is smaller than the full Hilbert space, and this restriction may result in a better-performing algorithm.

\subsubsection{Variational Hamiltonian ansatz}  Inspired by the QAOA ansatz, the variational Hamiltonian ansatz also aims to prepare a trial ground states for a given Hamiltonian $H=\sum_k H_k$ (where $H_K$ are Hermitian operators, usual Pauli strings) by Trotterizing an adiabatic state preparation process~\cite{wecker2015progress}. Here, each Trotter step corresponds to a variational ansatz so that  the unitary is given by $U(\vec{\theta})=\prod_l(\prod_k e^{-\theta_{l,k}H_k})$, and again is of the form Eq.~\eqref{eq:general-ansatz-chain}. Due to its versatility, the variational Hamiltonian ansatz has been  implemented for quantum chemistry~\cite{kivlichan2018quantum,wiersema2020exploring}, optimization~\cite{nasaQAOA2019}, and for quantum simulation problems~\cite{ho2019efficient}.

\subsubsection{Variable structure ansatz} In many ansatzes,  one optimizes over continuous parameters (such as rotation angles), while the structure of the circuit is kept fixed.  Although this enables the control of the overall circuit complexity, it may miss refinements attained by optimizing the circuit structure itself, including the addition or removal of unnecessary circuit elements. Optimizing the circuit structure was initially explored in a framework called ADAPT-VQE~\cite{grimsley2019adaptive}, which seeks to adaptively add specific elements to the ansatz to maximize the benefit while minimizing the number of circuit elements in quantum chemistry applications. (Improvements to ADAPT-VQE and variable ansatz for quantum chemistry have been introduced in Refs.~\cite{tang2019qubit,yordanov2020iterative}, and a variable structure version of the QAOA ansatz was introduced in Ref.~\cite{zhu2020adaptive}.) One can then view this problem as a sparse model problem, and whereas such an   optimization  is known to be hard, heuristic or greedy approximations that seek to add one term at a time have been shown to be helpful~\cite{grimsley2019adaptive,tang2019qubit}. 

Machine learning-aided evolutionary algorithms for circuit design have also been explored  in Refs.~\cite{rattew2019domain,chivilikhin2020mog}, where individuals (quantum circuits) from a population are upgraded to grow the circuit and explore the Hilbert space. In addition, Refs.~\cite{cincio2021machine,cincio2018learning,du2020quantum,zhang2020differentiable,bilkis2021semi} use tools from machine learning to develop variational ansatzes for various VQA applications. Complementary approaches based on exploring different ansatz variants simultaneously as an evolving cohort have also shown promising performance~\cite{rattew2020evoansatz}.

\subsubsection{Sub-logical ansatz and quantum optimal control} The parameters $\vec{\theta}$ are often specified at the logical circuit level (such as rotation angle), however sometimes they have a direct translation to device-level parameters below the logical level. Hence, one can include these device-level parameters in the definition of the ansatz, as this can offer additional flexibility~\cite{Yang2017optimizing}. This approach also establishes a connection to the idea of quantum optimal control, which is often used to determine the translation from logical to physical device parameters, and which is especially applicable for quantum simulations~\cite{magann2020pulses,choquette2020quantum}.  Refs.~\cite{Li2017hybrid,Lu2017} have explored using VQAs the construction of optimal control sequences. Although this can increase the number of parameters, the additional flexibility may allow for on-the-fly calibration effects that have been seen to reduce the effects of coherent noise~\cite{o2016scalable,Li2017hybrid,Lu2017}.

\subsubsection{ Hybrid ansatzes} In some cases, it is possible to combine quantum ansatzes with classical strategies to push some of the complexity onto the classical device.  For instance, in quantum chemistry one can exploit the classical simulability of free fermion dynamics to apply quantum operations via classical post-processing~\cite{PhysRevX.10.011004,Valiant2002quantum, Terhal2002classical, Jozsa2008matchgates,Mizukami2020orbital,Sokolov2020quantum}. A different approach is to use as ansatz a trainable linear combination of parametrized states $\ket{\psi(\{c_\mu\},\vec{\theta})} = \sum_\mu c_\mu \ket{\psi_\mu(\bm{\theta}_\mu)}$ with $\{c_\mu\}$ classically optimizable coefficients~\cite{mcclean2017hybrid, parrish2019quantum, parrish2019quantumfilter, Huggins2020nonorthogonal, stair2019multireference, bharti2020iterative,bharti2020quantum}. Moreover, given that quantum circuits can be viewed as tensor networks~\cite{Markov2008simulating}, it is natural to combine the existing tensor network techniques with a quantum ansatz~\cite{kim2017robust,kim2017holographic,Liu2019variational,barratt2020parallel,yuan2020quantum}. For instance, it has been shown that it is possible to unitarily contract tensor networks on a quantum computer~\cite{kim2017robust,kim2017holographic}. An alternative hybrid approach was proposed via the deep variational quantum eigensolver, where the algorithm divides the whole system into small subsystems and sequentially solves each subsystem and the interaction between the subsystems~\cite{fujii2020deep}. Finally, there is also a hybrid method that combines variational Monte Carlo techniques with a quantum ansatz to classically apply the so-called Jastrow operator $e^{\left(\sum_{i,j}J_{ij}\sigma_i \sigma_j\right)}$ (for $J$ a symmetric matrix, and $\sigma_i$ and $\sigma_j$ Pauli operators) to a parametrized quantum state $\ket{\psi(\vec{\theta})}$ with the goal of obtaining a more accurate result by optimizing together $J$ and $\vec{\theta}$ (Ref.~\cite{Mazzola2019nonunitary}).

\subsubsection{ Ansatz for mixed states} Since mixed states play an important role in many applications, such as systems at finite temperature, several ansatzes have been developed to construct a mixed state $\rho=\sum_{i}p_i\ket{\psi_i}\bra{\psi_i}$ of $n$ qubits (here $p_i$ are the eigenvalues of $\rho$ such that $\sum_i p_i=1$). A first approach (which comes at the cost of requiring up to $2n$ qubits)  is based on preparing a pure state that has  $\rho$ as a reduced state in some subsystem of qubits.   Refs.~\cite{Liu2019variational, Martyn2019product} have proposed a method to variationally obtain a purification $\ket{\psi}=\sum_i \sqrt{p_i}\ket{\psi_i}\ket{\phi_i}$ of $\rho$, whereas Ref.~\cite{yoshioka2019variational}  introduced a method to construct a state $\ket{\rho}=\frac{1}{c}\sum_{i}p_i \ket{\psi_i}\ket{\psi_i}$ with normalization $c=\sum_{i}p_i^2$. 
Alternatively, one can also train a probability distribution $\{p_i(\bm{\phi})\}$ and a set of states $\{\ket{\psi_i(\bm{\theta}_i)}\}$ to construct $\rho$ as the statistical ensemble $
\rho(\bm{\phi}, \{\bm{\theta}_i\}) = \sum_i p_i(\bm{\phi}) \ket{\psi_i(\bm{\theta}_i)}\bra{\psi_i(\bm{\theta}_i)}$. Ref.~\cite{Martyn2019product} proposed to use a simple product distribution based on physical insights, whereas a more general proposal for energy based models was introduced in Ref.~\cite{verdon2019quantumVQT}.  More recently, there has been a proposal to generate mixed states which uses the autoregressive model~\cite{Liu2020solving}.

\subsubsection{ Ansatz expressibility}  Given the wide range of ansatzes one can use, a relevant question is whether a given architecture can prepare a target state by optimizing its parameters. In this sense,  there are different ways to judge the quality of an ansatz~\cite{sim2019expressibility} by considering two different notions: the expressibility and the entangling capability of an ansatz. An ansatz is expressible if the circuit can be used to uniformly explore the entire space of quantum states. Thus one way to quantify the expressibility of an ansatz $U(\vec{\theta})$ is to compare the distribution of states obtained from $U(\vec{\theta})$ to the maximally expressive uniform (Haar) distribution of states $U_{\rm Haar}$. Motivated by this line of thought, the expressibility of a circuit is measured by~\cite{sim2019expressibility} $|| A^{(t)} ||$, where 
\begin{align}
    A^{(t)}(U) :=& \int d U_{\rm Haar } \, U_{\rm Haar }^{\otimes t}  | 0 \rangle \langle 0 | (U_{\rm Haar }^\dagger)^{\otimes t}\nonumber\\& -  \int d U \, U^{\otimes t}  | 0 \rangle \langle 0 | (U^\dagger)^{\otimes t} \, .
\end{align}
Other expressibility measures can be considered as well~\cite{sim2019expressibility}, and the expressibility of different ansatzes was investigated further in Ref.~\cite{nakaji2020expressibility}. Ref.~\cite{sim2019expressibility} also introduced a measure of entangling capability for ansatzes, which quantifies the average entanglement of states produced from randomly sampling the circuit parameters $\vec{\theta}$.
    
Quantifying expressibility for particular ansatzes is an active area of research~\cite{sim2019expressibility,nakaji2020expressibility,schuld2020effect,abbas2020power,holmes2021connecting}, with certain quantum architectures exhibiting higher expressibility (according to certain measures) relative to classical architectures~\cite{abbas2020power}.

\subsection{ Gradients}
\label{Sec2subSec:Gradients}

Once the cost function and ansatz have been defined, the next step is to train the parameters $\vec{\theta}$ and solve the optimization problem of Eq.~\eqref{eq:optimization}. It is known that for many optimization tasks using information in the cost function gradient (or in higher-order derivatives) can help in speeding up and guaranteeing the convergence of the optimizer. One of the main advantages of many VQAs is that, as discussed below, one  can analytically evaluate the cost function gradient. 
    
\subsubsection{ Parameter-shift rule} Let us consider for simplicity a cost function of the form in Eq.~\eqref{eq:cost-simple} with $f_k(x)=x$, and let  $\theta_l$ be the $l$-th element in $ \vec{\theta}$ which parametrize a unitary $e^{i\theta_l\sigma_{l}}$ in the ansatz. Here, $\sigma_l$ is a Pauli operator. Surprisingly, there is a hardware-friendly protocol to evaluate the partial derivative of $C(\vec{\theta})$ with respect to $\theta_l$ often referred to as the parameter-shift rule~\cite{guerreschi2017practical, mitarai2018quantum,schuld2019evaluating,bergholm2018pennylane}. Explicitly, the parameter-shift rules states that the equality 
\begin{align}\label{eq:parameter-shift}
    \frac{\partial C}{\partial \theta_l} = \sum_{k}\frac{1}{2\sin \alpha}&\Big( \Tr\left[O_{k} U\ad(\vec{\theta}_+) \rho_k U(\vec{\theta}_+)\right]\nonumber\\& -\Tr\left[O_{k} U\ad(\vec{\theta}_-) \rho_k U(\vec{\theta}_-)\right]\Big)\,,
\end{align}
with $\vec{\theta}_{\pm}=\vec{\theta}\pm\alpha\vec{e}_l$, holds for any real number $\alpha$. Here  $\vec{e}_l$ is a vector having $1$ as its $l$-th element and $0$ otherwise. Equation~\eqref{eq:parameter-shift} shows that one can evaluate the gradient by shifting the $l$-th parameter by some amount $\alpha$. Note that the accuracy of the evaluation depends on the coefficient $1/(2\sin\alpha)$ since each of the $\pm \alpha$-term is evaluated by sampling $O_k$. This accuracy is maximized at $\alpha=\pi/4$, since $1/\sin\alpha$ is minimized at this point. Although the parameter-shift rule might resemble a naive finite difference, it evaluates the analytic gradient of the parameter by virtue of the coefficient $1/\sin\alpha$. A detailed comparison between the parameter-shift rule and the finite difference can be found in Ref.~\cite{mari2020estimating}. Finally, the gradient for more general $f_k(x)$ can be obtained from Eq.~\eqref{eq:parameter-shift} by using the chain rule.

\subsubsection{ Other derivatives}  Higher-order derivatives of the cost function can be evaluated by straight-forward extensions of the parameter-shift rule.     For example, the second derivative for the previous example can be written as
\small
\begin{align}
    \frac{\partial^2 C}{\partial \theta_l^2} = \sum_k\frac{1}{4\sin^2\alpha}&\Big( \Tr\left[O_k U\ad\left(\vec{\theta}+2\alpha\vec{e}_l\right)\rho_k U\left(\vec{\theta}+2\alpha\vec{e}_l\right)\right]\nonumber \\
    & + \Tr\left[O_k U\ad\left(\vec{\theta}-2\alpha\vec{e}_l\right) \rho_k U\left(\vec{\theta}-2\alpha\vec{e}_l\right)\right]\nonumber\\
    &- 2\Tr\left[O_k U\ad(\vec{\theta}) \rho_k U(\vec{\theta})\right]\Big)\,,\nonumber
\end{align}
\normalsize
by applying the parameter-shift rule twice.
Other higher-order ones such as $\frac{\partial^2 C}{\partial \theta_l\theta_{l'}}$ or $\frac{\partial^3 C}{\partial \theta_l^3}$ can be obtained in a similar fashion. Explicit formulas can be found in Refs.~\cite{cerezo2020impact,mari2020estimating}. These observations relate to the fact that the cost function can be expanded into a trigonometric series that admits a classically efficient, analytical approximation around any reference point. One can thus infer a classical model of the cost function, and minimise it, to offload more work from the quantum processor to the classical supervising system~\cite{nakanishi2019,koczor2020qad}.

Other types of derivatives of the parametrized quantum state not directly related to the cost function, such as a metric tensor of a state $\frac{\partial \bra{\psi(\vec{\theta})}}{\partial \theta_l}\frac{\partial \ket{\psi(\vec{\theta})}}{\partial \theta_l'}$ (with $\ket{\psi(\vec{\theta})}=U(\vec{\theta})\ket{\psi_0}$ for some initial state $\ket{\psi_0}$), are sometimes used in sophisticated optimization algorithms~\cite{mcardle2019variational, stokes2019quantum, koczor2019quantum}  and variational quantum simulation \cite{li2017efficient,yuan2019theory,endo2018variational} (see the section on dynamical quantum simulation). 
As quantities such as $\frac{\partial \bra{\psi(\vec{\theta})}}{\partial \theta_l}\frac{\partial \ket{\psi(\vec{\theta})}}{\partial \theta_l'}$ are essentially overlaps of different states, this can be evaluated via Hadamard-test like protocols~\cite{yuan2019theory}.
However, as shown in Ref.~\cite{mitarai2019methodology}, those can also be reduced to the parameter-shift technique.

\subsection{Optimizers}\label{sec:optimizers}

As for any variational approach, the success of a variational quantum algorithm (VQA) depends on the efficiency and reliability of the optimization method used. The classical optimization problems associated with VQAs are expected to be NP-hard in general as they involve cost functions that can have many local minima~\cite{bittel2021training}. In addition to the typical difficulties encountered in complex classical optimizations, it has been shown that when training a VQA one can encounter new challenges. These include issues such as the inherently stochastic environment due to the finite budget for measurements, hardware noise, and the presence of barren plateaus (see main text). This has led to the development of many quantum-aware optimizers, with the optimal choice still being an  active topic of debate. Here we discuss a selection of optimizers that have been designed or promoted for use with VQAs. For convenience, these will be grouped into two categories based on whether or not they implement some version of gradient descent. 

\subsubsection{Gradient descent methods} One of the most common approaches to optimization is to make iterative steps in directions indicated by the gradient. Given that only statistical estimates are available for these gradients, these strategies fall under the umbrella of Stochastic Gradient Descent (SGD). One SGD method that has been imported from the machine learning community is Adam, which adapts the size of the steps taken during the optimization to allow for more efficient and precise solutions than those obtained through basic SGD~\cite{Kingma2015}. An alternative method inspired by the machine learning literature  adapts the precision (the number of shots taken for each estimate), rather than the step size, at each iteration in an attempt to be frugal with the quantum resources used~\cite{kubler2019adaptive}. It is possible to attain an unbiased estimator for a partial derivative with even just a single shot~\cite{sweke2019stochastic}, so adapting the number of shots when low precision is acceptable can lead to significant reductions in the overall shot cost of an algorithm.
    
A different gradient-based approach is based on simulating an imaginary time evolution~\cite{mcardle2019variational}, or equivalently by using the quantum natural gradient descent method, which is based on notions of information geometry~\cite{stokes2019quantum,koczor2019quantum}. Whereas standard gradient descent takes steps in the steepest descent direction in the $l_2$ (Euclidean) geometry of the parameter space, natural gradient descent works instead on a space with a metric tensor that encodes the sensitivity of the quantum state to variations in the parameters. Using this metric tensor, typically accelerates the convergence of the gradient update steps, allowing a given level of precision to be attained with fewer iterations. This method has also been extended to incorporate the effects of noise~\cite{koczor2019quantum}.

\subsubsection{Other methods} A different method which uses gradients, but has a more complicated update step than SGD, is meta-learning~\cite{wilson2019optimizing}. In this context, the optimizer `learns to learn' by training a neural network to make a good update step based on the optimization history and current gradient with similar optimization problems. Because the update steps taken are based on rules learned from similar cost functions, this meta-learning approach has significant potential to be highly efficient when used on a new instance of a common class of optimizations.
    
Of the optimization methods proposed for use with VQAs which do not directly utilize gradients, the one that is perhaps the most closely related to SGD is the simultaneous perturbation stochastic approximation (SPSA) method~\cite{spall1992}. SPSA can be considered as an approximation to gradient descent where the gradient is approximated by a single partial derivative computed by a finite difference along a randomly chosen direction. SPSA has thus been put forward as an efficient method for VQAs as it avoids the expense of computing many gradient components at each iteration. Moreover, it has been shown that for a restricted set of problems,  SPSA has a faster theoretical convergence rate (in terms of the number of function calls) than SGD performed with finite differences~\cite{spall1992}.
    
Finally, another noteworthy gradient-free approach has been developed specifically for the context of VQAs for problems where the objective function is a linear function of an operator expectation value, so that $C(\vec{\theta})$ can be expressed as a sum of trigonometric functions. Using this insight, one can fit the functional dependence on a few parameters (with the rest held fixed) allowing one to make local parameter updates~\cite{nakanishi2019,parrish2019}. Performing such local updates sequentially over all parameters, or subsets of parameters, and iterating over all parameters one then has an optimization method that is gradient-free and which does not depend on hyper-parameters. Additionally, a variation of this method using Anderson acceleration (a method that adds a linear combination of prior steps to each new update step) to speed up convergence has been proposed~\cite{parrish2019}. 

\subsubsection{Convergence analysis} The cost landscapes of VQAs are generally non-convex and can be complicated~\cite{huembeli2020characterizing}, making it difficult to obtain general guarantees about the computational expense of the optimizations. However, for simplified landscapes, SGD convergence guarantees have been derived which are similar to those provided in the machine learning literature~\cite{harrow2019low,sweke2019stochastic}. Furthermore, within a convex region about a minimum, SGD methods using gradients calculated via the parameter-shift rule have been shown to have smaller upper bounds on the optimization complexity than methods using only objective values (including finite difference methods)~\cite{harrow2019low}.

\begin{figure}[t] 
    \centering
    \includegraphics[width=1\columnwidth]{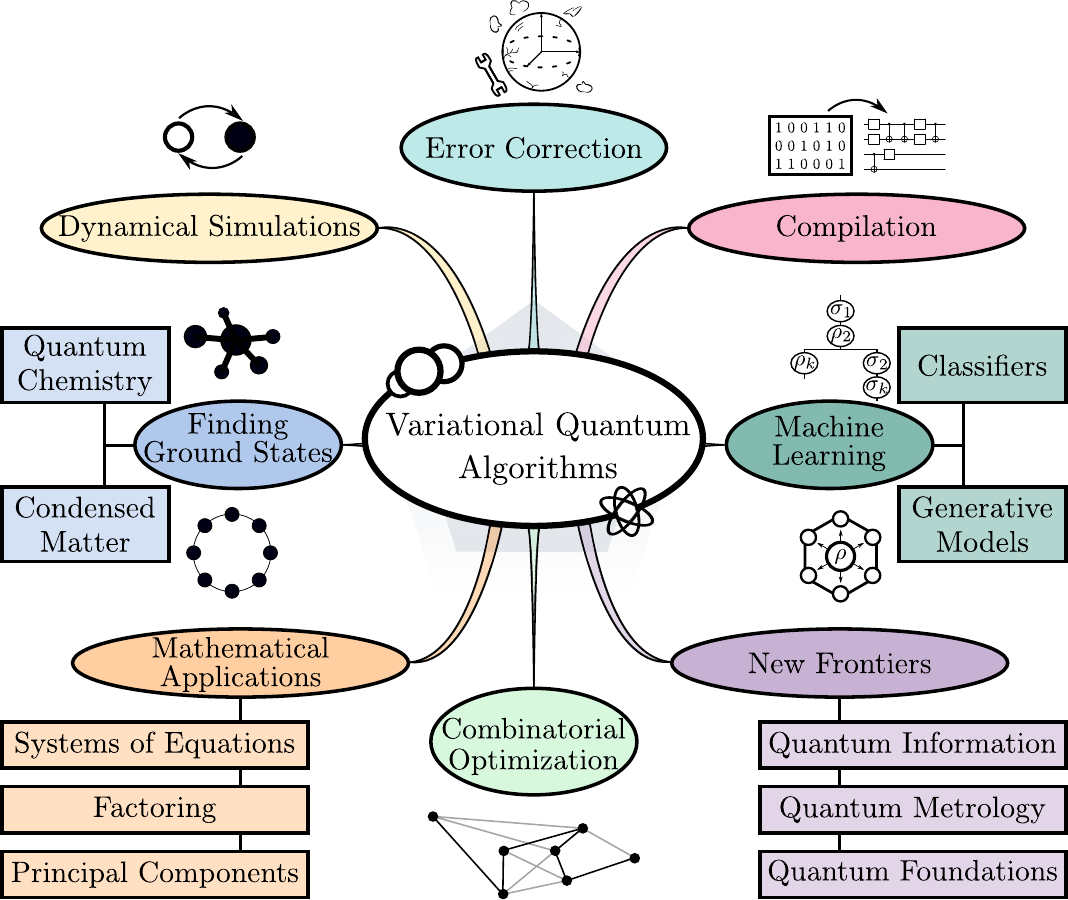}
    \caption{\textbf{Applications of Variational Quantum Algorithms (VQAs)}. Many applications have been envisioned for VQAs. Here we show some of the key applications that are discussed in this Review.}
    \label{fig:Appli}
\end{figure}

\section{ Applications}\label{sec:3}

One of the main advantages of the VQA paradigm is that it allows for task-oriented programming. That is, VQAs provide a framework that can be used to tackle a wide array of tasks. This has lead to VQAs being proposed for essentially all applications envisioned for quantum computers, and in fact, it has been shown that VQAs allow for universal quantum computing~\cite{biamonte2019universal}.  In this section we provide an overview of some of the main applications of VQAs and their state-of-the implementation.  These applications are also summarized in Figure~\ref{fig:Appli}. We also refer the reader to Section~\ref{sec:5} for an overview of applications where VQAs can be potentially used to obtain a quantum advantage.

\subsection{ Finding ground and excited states}\label{sec:ground-state}

The best-known application of VQAs is estimating low-lying eigenstates and corresponding eigenvalues of a given Hamiltonian. Previous quantum algorithms to find the ground state of a given Hamiltonian $H$ were based on adiabatic state preparation and quantum phase estimation subroutines~\cite{QPEGroundState99,aspuru2005simulated}, both of which have circuit depth requirements beyond those available in the NISQ era.  Hence, the first proposed VQA, the Variational Quantum Eigensolver (VQE), was developed to provide a near-term solution to this task.  Here we review both the original VQE architecture and some more advanced methods for finding ground and excited states. 

\begin{figure}[t] 
    \centering
    \includegraphics[width=.8\columnwidth]{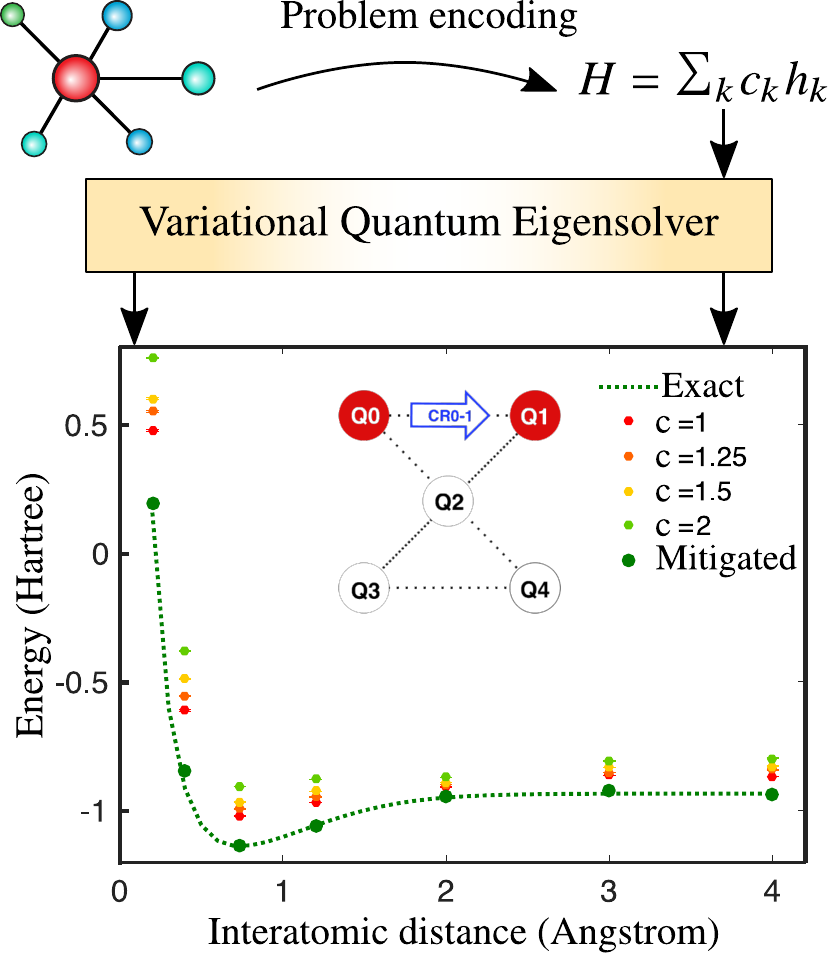}
    \caption{\textbf{Variational Quantum Eigensolver (VQE) implementation}. The VQE algorithm can be used to estimate the ground state energy $E_G$ of a molecule. The interactions of the system are encoded in a Hamiltonian $H$, usually expressed as a linear combination of simple operators $h_k$ with coefficients $c_k$. Taking $H$ as input, VQE outputs an estimate $\widetilde{E}_G$ of the ground-state energy. The lower part of the figure shows the results of a VQE implementation for the electronic structure problem of an $H_2$ molecule, whose exact energy is shown as a dashed line. The experimental results were obtained using two of the five qubits in one of IBM's superconducting quantum processors (the inset illustrates qubit connectivity with $Q_0 \dots Q_4$ denoting the qubits ). Due to the presence of hardware noise the estimated energy $\widetilde{E}_G$ has a gap with the true energy. In fact, amplifying the noise strength (that is increasing the quantity $s$), deteriorates the solution quality. However, as discussed below, one can use error mitigation techniques to improve the solution quality. Figure adapted from Ref.~\cite{kandala2019error}, Springer Nature Limited.  }
    \label{fig:VQE}
\end{figure}

\subsubsection{ Variational quantum eigensolver}  As shown in Fig.~\ref{fig:VQE}, VQE  is aimed at finding the ground state energy $E_G$ of a Hamiltonian $H$~\cite{VQE}. Here the cost function is defined as $C(\vec{\theta})=\bra{\psi(\vec{\theta})}H \ket{\psi(\vec{\theta})}$. That is, one seeks to minimize the expectation value of $H$ over a trial state $\ket{\psi(\vec{\theta})}=U(\vec{\theta})\ket{\psi_0}$ for some ansatz $U(\vec{\theta})$ and initial state $\ket{\psi_0}$. According to the Rayleigh-Ritz variational principle, the cost is meaningful and faithful as $C(\vec{\theta})\geq E_G$, with the equality holding if $\ket{\psi(\vec{\theta})}$ is the ground state $\ket{\psi_G}$ of $H$. In practice, the Hamiltonian $H$ is usually represented as a linear combination of products of Pauli operators $\sigma_k$ as $H=\sum_k c_k \sigma_k~(c_k \in \mathbb{R})$, so that the cost function $C(\vec{\theta})$ is obtained from a linear combination of expectation values of $\sigma_k$. Since practical physical systems are generally described by sparse Hamiltonians, the cost function can be efficiently estimated on quantum computers with a computational cost that usually grows at most polynomially with the system size.

\subsubsection{ Orthogonality constrained VQE} Once an approximated ground state $\ket{\tilde{\psi}_G}=U(\vec{\theta}^*)\ket{\psi_0}$ has been obtained, one can use it to find excited states of $H$. Let $a$ be a positive constant that is much larger than the energy gap between the ground state and the first excited states. Then, $H'=H+a \ket{\tilde{\psi}_G}\bra{\tilde{\psi}_G}$ is a Hamiltonian whose ground state is the first excited state of $H$ (Ref.~\cite{higgott2019variational}). Thus, by using the VQE for $H'$ with an updated cost  $C(\vec{\theta})=\bra{\psi(\vec{\theta})}H \ket{\psi(\vec{\theta})}+a \langle \psi(\vec{\theta})| \tilde{\psi}_G\rangle \langle \tilde{\psi}_G|\psi(\vec{\theta}) \rangle $, one may find the first excited state of $H$. The first term here is evaluated as in VQE, and the second term can be obtained by computing the state overlap between $| \tilde{\psi}_G\rangle$ and $|\psi(\vec{\theta}) \rangle$ (Refs.~\cite{buhrman2001quantum,cincio2018learning}). This procedure can be further generalized to approximate higher excited states. Moreover, it has been shown that incorporating an imaginary time evolution can help to improve the calculation robustness~\cite{jones2019variational}.

\subsubsection{ Subspace expansion method}    Another way to discover low energy excited states using information of the estimated ground state $\ket{\tilde{\psi}_G}$ is via the subspace expansion method~\cite{mcclean2017hybrid}. Here one runs an additional optimization in a  subspace of states $\{ \ket{\psi_k} \}$ generated from $\ket{\tilde{\psi}_G}$. For instance, one creates states $\ket{\psi_k} = \sigma_k \ket{\tilde{\psi}_G}$ for low-weight Pauli operators $\sigma_k$, and expands the candidates for the eigenstates as $\ket{E}=\sum_k \alpha_k \ket{\psi_k}$. Then, one obtains approximations to the lowest eigenstates by training the  coefficients $\vec{\alpha}=(\alpha_0, \alpha_1, \alpha_2,...)$ while solving the generalised eigenvalue problem $H \vec{\alpha}= E S \vec{\alpha}$, with $H_{k,j}= \bra{\psi_k} H \ket{\psi_j}$ and $S_{k,j}=\langle \psi_k| \psi_j \rangle$.

\subsubsection{ Subspace VQE} The main idea behind subspace VQE~\cite{nakanishi2019subspace} is to train a unitary for preparing states in the lowest energy subspace of $H$. There are two variants of subspace VQE called weighted and non-weighted subspace VQE. 
For the weighted subspace VQE, one considers a cost function 
$C(\vec{\theta})=\sum_{i=0}^m w_i \bra{\varphi_i}U(\vec{\theta}) H U (\vec{\theta}) \ket{\varphi_i}$ with ordered weights $w_0>w_1> \cdots > w_m$ and easily prepared mutually-orthogonal states $\{\ket{\varphi_i}\}$. By minimizing the cost function, one approximates the subspace of the lowest eigenstates as $\{ U(\vec{\theta}^*) \ket{\varphi_i}\}_{i=0}^m$. Since the weights are in decreasing order, each state $U(\vec{\theta}^*) \ket{\varphi_i}$ corresponds to an eigenstate of the (non-degenerate) Hamiltonian with increasing energies. 

The non-weighted subspace VQE makes use of the cost function $C_1(\vec{\theta})=\sum_{i=0}^m \bra{\varphi_i}U\ad(\vec{\theta}) H U (\vec{\theta}) \ket{\varphi_i}$. Minimizing $C_1$ again gives the subspace of lowest eigenstates. As each state $U(\vec{\theta}^*) \ket{\varphi_i}$ is in a superposition of the eigenstates, one needs to  further optimize a  second cost $C_2(\vec{\theta}^*,\vec{\phi})=\bra{\varphi_i} V\ad(\vec{\phi}) U\ad(\vec{\theta}^*) H U (\vec{\theta}^*) V(\vec{\phi}) \ket{\varphi_i}$ over parameters $\vec \phi$ to rotate each state $U (\vec{\theta}^*) V(\vec{\phi}) \ket{\varphi_i}$ to an eigenstate.

\subsubsection{ Multistate contracted VQE} The multistate contracted VQE~\cite{parrish2019quantum} can be regarded as a midway point between subspace expansion and subspace VQE. It first obtains the lowest energy subspace $\{ U(\vec{\theta}^*) \ket{\varphi_i}\}_{i=0}^m$ by optimizing  $C_1(\vec{\theta})$ as in the non-weighted subspace VQE. Instead of optimizing an additional unitary, the multistate contracted VQE  approximates each eigenstate as $\ket{E}=\sum_i \alpha_i U(\vec{\theta}^*) \ket{\varphi_i}$ with coefficients $\alpha_i$ which are obtained by solving a generalised eigenvalue problem similar to subspace expansion with $S=\id$.

\subsubsection{ Adiabatically assisted VQE} Quantum adiabatic optimization seeks to find a solution to an optimization problem by slowly transforming the ground state of a simple problem to that of a complex problem. These methods have a close connection with classical homotopy schemes that are used to find the solutions of classical problems in optimization~\cite{mcclean2020low}.  In light of this connection, the adiabatically assisted VQE~\cite{garcia2018addressing} uses a cost function $C(\vec{\theta})=\bra{\psi(\vec{\theta})}H(s) \ket{\psi(\vec{\theta})}$, where $H(s)=(1-s)H_0+sH_P$ and $\ket{\psi(\vec{\theta})}=U(\vec{\theta})\ket{\psi_0}$.  Here  $H_P$ is the problem Hamiltonian of interest and $H_0$ is a simple Hamiltonian whose known ground state is taken as the initial state $\ket{\psi_0}$. During the parameter optimization, one slowly changes $s$ from $0$ to $1$.  The idea of Hamiltonian transformation has been used as a type of ansatz to obtain solutions near the more challenging endpoint~\cite{cerezo2020variational}.

\subsubsection{ Accelerated VQE} As previously mentioned, whereas Quantum Phase Estimation (QPE) provides a means to estimate eigenenergies in the fault-tolerant era, it is not implementable in the near-term. However, one of the positive features of this algorithm is that a precision $\epsilon$ can be obtained with a number of measurements which scale as $\OC(\log (\frac{1}{\epsilon}))$. This is in contrast with VQE, which requires $\OC(\frac{1}{\epsilon^2})$ measurement for the same precision.  This scaling motivated the Accelerated VQE algorithm, which interpolates between the VQE and QPE algorithms~\cite{wang2019accelerated,wang2020bayesian,wang2020bayesian2}. The interpolation involves taking the VQE algorithm and replacing the measurement process with a tunable version of QPE called $\alpha$-QPE. This allows the measurement cost to interpolate between that of VQE and QPE.

\subsection*{Dynamical quantum simulation}\label{sec:dynamical-quantum-simulation}
Apart from static eigenstate problems, VQAs can also be applied to simulate the dynamical evolution of a quantum system. Conventional quantum Hamiltonian simulation algorithms, such as the Trotter-Suzuki product formula~\cite{nielsen_chuang}, generally discretize time into small time steps and simulate each time evolution with a quantum circuit. Therefore, the circuit depth generally increases polynomially with the system size and simulated time.
Given the noise inherent in NISQ devices, the accumulated hardware errors for such deep quantum circuits can prove prohibitive. To address this, VQAs for dynamical quantum simulation only use a shallow depth circuit, significantly reducing the impact of hardware noise.

\subsubsection{ Iterative approach} Instead of directly implementing the unitary evolution described by the Schr\"odinger equation $\frac{d\ket{\psi(t)}}{dt} =-iH \ket{\psi(t)}$, iterative variational algorithms~\cite{li2017efficient,yuan2019theory} consider trial states $\ket{\psi(\vec \theta)}$ and map the evolution of the state to the evolution of the parameters $\vec \theta$. By iteratively updating the parameters, the quantum state is effectively updated and hence evolved. Specifically, by using variational principles, such as McLachlan's principle~\cite{mclachlan1964variational} to solve the minimization $\mathrm{min}_{\dot{\vec{\theta}}} \delta \|(\frac{d}{dt}+iH) \ket{\psi(\vec{\theta})} \|$, one obtains a linear equation for the parameters as $M \cdot \dot{\vec{\theta}} = V$. Here $\|\ket{\psi}\|=\sqrt{\bra{\psi}{\psi}\rangle}$, $\dot{\vec\theta} = \frac{d \vec\theta}{dt}$, $M_{i,j}=\mathrm{Re}\big(\partial_i\bra{\psi(\vec{\theta})}  \partial_j \ket{\psi(\vec{\theta})} \big)$, $V_i=\mathrm{Im}\big(\bra{\psi(\vec{\theta})} H \partial_i \ket{\psi(\vec{\theta})} \big)$, and $ \partial_i \ket{\psi(\vec{\theta})}=\frac{\partial \ket{\psi(\vec{\theta})}}{\partial\vec{\theta}_i}$. Each element of $M$ and $V$ can be efficiently measured with a modified  Hadamard test circuit. By solving the linear equation, one can iteratively update the parameters from $\vec{\theta}$ to $\vec{\theta}+ \dot{\vec\theta} \Delta t $ with a small time step $\Delta t$. 
Similar variational algorithms could be applied for simulating the Wick-rotated Schr\"odinger equation of imaginary time evolution~\cite{mcardle2019variational} and general first-order derivative equations with non-Hermitian Hamiltonians~\cite{endo2018variational}. A systematic comparison between different variational principles for different problems can be found in Ref.~\cite{yuan2019theory}. Recent works also extend the algorithms to use adaptive ansatz to reduce the circuit depth~\cite{yao2020adaptive,zhang2020lowdepth}

\subsubsection{ Subspace approach} The weighted subspace VQE~\cite{nakanishi2019subspace} provides an alternative way to simulate dynamics in the subspace of the low energy eigenstates~\cite{heya2019subspace}. 
Here one uses the weighted subspace VQE unitary operator $U(\vec{\theta}^*)$ that maps computational basis states $\{\ket{\varphi_j}\}$ to the low energy eigenstates $\{\ket{E_j}\}$ as $U(\vec{\theta}^*) \ket{\varphi_j}\approx e^{i \delta_j} \ket{E_j}$, with $\delta_j$ an unknown phase. 
Considering the low energy subspace, the time evolution operator can be approximated as $\mathrm{exp}(-i H t) \approx U(\vec{\theta}^*) \mathcal T(t) U^\dag(\vec{\theta}^*)$ with $\mathcal T(t)=\sum_{j} \exp(-iE_j t)\ket{\psi_j}\bra{\psi_j}$. The procedure could intuitively be understood as first, rotating the state to the computational basis with $U^\dag(\vec{\theta}^*)$, second, evolving  the state with $\mathcal T(t)$, and third, rotating the basis back with $U(\vec{\theta}^*)$. 
Therefore, for any state $\ket{\psi(0)}=\sum_j \alpha_j \ket{E_j}$ that is a superposition of the low energy eigenstates, its time evolution can be simulated as $\ket{\psi(t)}=U(\vec{\theta}^*) \mathcal T(t) U^\dag(\vec{\theta}^*)\ket{\psi(0)}$. Since the time evolution is directly implemented via $\mathcal T(t)$, it does not involve iterative parameter update and the circuit depth is independent of the simulation time.

\subsubsection{ Variational fast forwarding} Similar to the subspace approach, variational fast forwarding~\cite{cirstoiu2019variational,gibbs2021long} simulates the time evolution operation $\mathrm{exp}(-i H t)$ as $U(\vec{\theta}^*) \mathcal T(\vec E, t) U^\dag(\vec{\theta}^*)$ with  $\mathcal T(\vec E, t)=\sum_{j} \exp(-iE_j t)\ket{\psi_j}\bra{\psi_j}$ a trainable diagonal matrix  and  $U(\vec{\theta}^*)$ a trainable unitary that maps between the eigenstates of $H$ and the computational basis. Although the subspace approach obtains $\mathcal T(\vec E, t)$ and $U(\vec{\theta}^*)$ via weighted subspace VQE, variational fast forwarding optimises a cost given by the fidelity between $e^{-iH \delta t}$ and $U(\vec{\theta}^*) \mathcal T(\vec E, \delta t) U^\dag(\vec{\theta}^*)$ for a small time step $\delta t$ via the so-called local Hilbert-Schmidt test~\cite{QAQC}. 
Then, according to the Trotter-Suzuki product formula, one has $e^{-iH T} = (e^{-iH \Delta t})^M \approx U(\vec{\theta}^*)( \mathcal T(\vec E, t))^M U^\dag(\vec{\theta}^*)$. Again, since the time evolution is implemented in $\mathcal T(\vec E, t)$, one can simulate the evolution for arbitrary time $t$ with the same circuit structure. As shown in Ref.~\cite{commeau2020variational}, the ensuing Trotter error of this approach can be removed by diagonalizing instead the Hamiltonian $H$ that generates the evolution.

\subsubsection{ Simulating open systems} The VQA framework can also be extended to simulate dynamical evolution of open quantum systems. Suppose that the dynamics of the system is described by $\frac{d \rho}{dt} = \mathcal{L}(\rho)$, where $\mathcal{L}$ denotes a super-operator for a dissipative process. Similarly to the iterative approach for pure states~\cite{li2017efficient}, one maps the evolution of the mixed state to one of the variational parameters via McLachlan's principle, which solves the minimization $\min_{\dot{\vec \theta}}\|(\frac{d}{dt}-\mathcal{L})\rho(\vec\theta)\|$. The solution determines the evolution of the parameters $ M\cdot \dot{\vec{\theta}}= V$   with  $M_{i,j}=\mathrm{Tr}\big[\partial_i \rho(\vec{\theta})^\dag \partial_j  \rho(\vec{\theta}) \big]$, $V_i=\mathrm{Tr}\big[\partial_i \rho(\vec{\theta})^\dag \mathcal{L}(\rho) \big]$ and $\partial_i \rho(\vec{\theta})= \frac{\partial \rho(\vec{\theta})}{\partial \vec{\theta}_i}$. Each term of $M$ and $V$ can be computed by applying the SWAP test circuit on two copies of the purified states~\cite{yuan2019theory}. Here, to simulate an open system of $n$ qubits, one needs to apply operations on $4n+1$ qubits. An alternative approach~\cite{endo2018variational} which reduces this overhead is to simulate the stochastic Schr\"odinger equation, which unravels the evolution of the density matrix into trajectories of pure states. Each pure state trajectory experiences continuous damping effect and jump processes due to the noise operators, both of which can be efficiently simulated. Since this method one only controls a single copy of the pure state, it only requires $n+1$ qubits.

\subsection{ Optimization}\label{sec:optimization}

\begin{figure}[t] 
    \centering
    \includegraphics[width=.8\columnwidth]{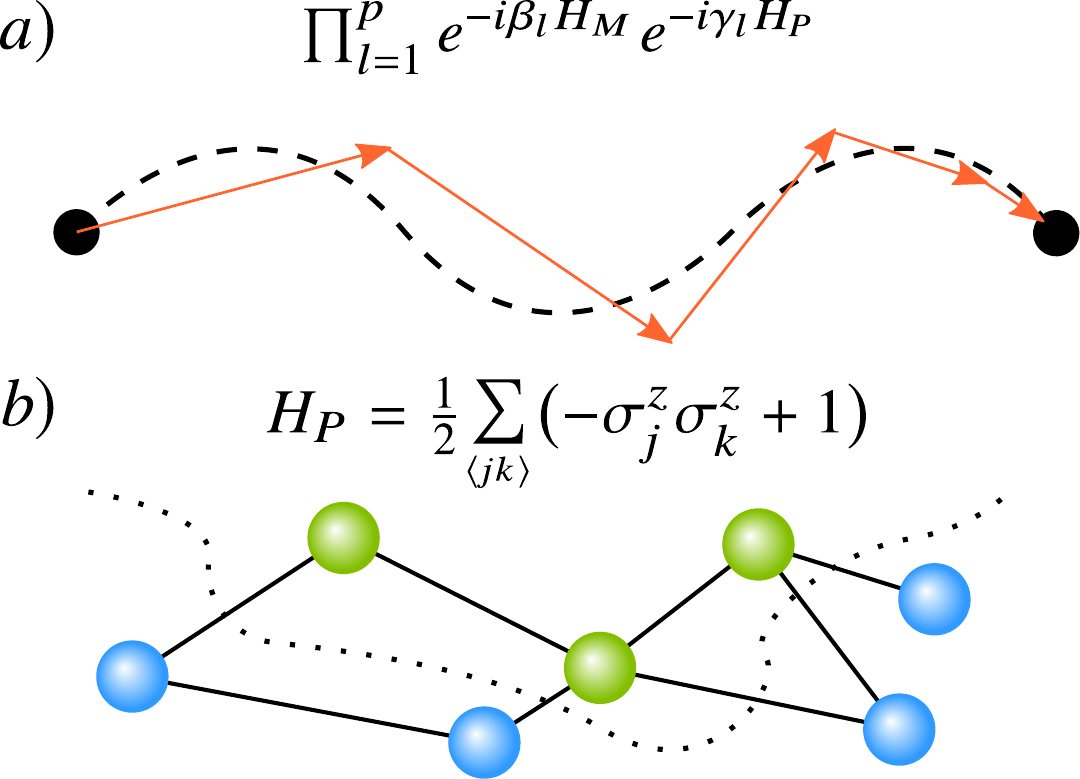}
    \caption{\textbf{Quantum Approximate Optimization Algorithm (QAOA)}. a. Schematic representation of the Trotterized adiabatic transformation in the ansatz. The algorithm only loosely follows the evolution of the ground state of $H(t) = (1-t)H_M + tH_P$ for every $t \in [0,1]$, as one is interested in making the final state close to the ground state of the problem Hamiltonian $H_P$, with $H_M$ being a mixer Hamiltonian. The free parameters $\{\beta_l\}_{l=1}^p$ and $\{\gamma_l\}_{l=1}^p$ are trained, with $p$ being the number of QAOA rounds. b. Problem Hamiltonian $H_P$ and graph $\langle j k \rangle$ for a Max-Cut task. Each node in the graph (circle) represents a spin. Vertices connecting  two nodes indicate an interaction $\sigma^z_j\sigma^z_k$ in  $H_P$, with $\sigma^z_k$ the Pauli $z$ operator on spin $k$. The solution is encoded in the ground state of $H_P$ where some spins are pointing up (green) whereas others point down (blue).
    }
    \label{fig:QAOA}
\end{figure}

Thus far we have discussed using VQAs for tasks which are inherently quantum in nature, that is, finding ground states and simulating the evolution of quantum states. In this subsection we discuss a different possibility where one uses a VQA to solve a classical optimization problem~\cite{moll2018quantum}. 

The most famous VQA for quantum-enhanced optimization is the QAOA~\cite{qaoa2014}, originally introduced to approximately solve combinatorial problems such as Constraint-Satisfaction (SAT)~\cite{Lin2016sat} and Max-Cut problems~\cite{qaoaMaxCut2018}.

Combinatorial optimization problems are defined on binary strings $\vec{s}=(s_1,\cdots,s_N)$ with the task of maximizing a given classical objective function $L(\vec{s})$. QAOA encodes $L(\vec{s})$ in a quantum Hamiltonian $H_P$ by promoting each classical variable $s_j$ to a Pauli spin-$1/2$ operator $\sigma^z_j$, so that the goal is to prepare the ground state of $H_P$. Motivated by the quantum adiabatic algorithm, QAOA replaces adiabatic evolution with $p$ rounds of alternating time propagation between the problem Hamiltonian $H_P$ and appropriately chosen mixer Hamiltonian $H_M$, see Fig.~\ref{fig:QAOA}. As discussed in the subsection on quantum alternating operator ansatz, the evolution time intervals are treated as variational parameters and are optimized classically. Hence, defining $\vec{\theta}=\{\vec{\gamma},\vec{\beta}\}$, the cost function is $C(\vec{\gamma},\vec{\beta})=\bra{\psi_p(\vec{\gamma},\vec{\beta})} H_P \ket{\psi_p(\vec{\gamma},\vec{\beta})}$ with
\small
\begin{equation} \label{eq:QAOAansatz}
\ket{\psi_p(\vec{\gamma},\vec{\beta})} = 
e^{-i \beta_p H_M} e^{-i \gamma_p H_P} \cdots
e^{-i \beta_1 H_M} e^{-i \gamma_1 H_P} \ket{\psi_0} \,,
\end{equation}
\normalsize
and where $\ket{\psi_0}$ is the ground state of $H_M$.

Finding optimal values $\vec{\gamma}$ and $\vec{\beta}$ is a hard problem since the optimization landscape in QAOA is non-convex with many local optima~\cite{shaydulin2019multistart}. Hence, great efforts have been devoted to finding a good classical optimizer that would require as few calls to the quantum computer as possible. Gradient-based~\cite{romero2018strategies, crooks2018performance}, derivative-free~\cite{wecker2016training, Yang2017optimizing}, and reinforcement learning~\cite{khairy2020learning} methods were investigated, and this still remains an active field to guarantee a good performance for the QAOA. 

\subsection{ Mathematical applications}

Several VQAs have been proposed to tackle relevant mathematical problems such as solving linear systems of equations or integer factorization. Since in many cases there exist quantum algorithms for the fault-tolerant era aimed for these tasks, the goal of VQAs is to have heuristical scalings comparable to the provable scaling of these non-near-term algorithms while keeping the algorithm requirements compatible with the NISQ era.   

\subsubsection{ Linear systems} Solving systems of linear equations has wide-ranging applications in science and engineering. 
Quantum computers offer the possibility of exponential speedup for this task. Specifically, for an $N\times N$ linear system $A\vec{x} =\vec{b}$ (with $A$  an $N\times N$ matrix, and $\vec{b}$ an $N\times 1$ column vector defined from the linear systems problem), one considers the Quantum Linear Systems Problem (QLSP) where the task is to prepare a normalized state $\ket{x}$ such that $A\ket{x} \propto \ket{b}$, where $\ket{b}=\vec{b}/\|\vec{b}\|$ is also a normalized state. The classical algorithmic complexity for this task scales polynomially in the dimension $N$, whereas the now-famous Harrow--Hassidim--Lloyd (HHL) quantum algorithm~\cite{harrow2009quantum} has a complexity that scales logarithmically in $N$, with some scaling improvements having been proposed~\cite{Ambainis,Yigit,CKS,Chakraborty}. These pioneering quantum algorithms, however, will be difficult to implement in the near-term due to the enormous circuit depth requirements~\cite{scherer2017concrete}. 

This situation has motivated VQAs for the QLSP~\cite{bravo-prieto2019, Xiaosi, huang2019near}. A common feature in these algorithms is the assumption that $A = \sum_k c_k A_k$ is given as a linear combination of unitaries $A_k$ that can be efficiently implemented weighted by real coefficients $c_k$. One can then construct a Hamiltonian whose ground state is the solution to the QLSP and apply a variational approach to minimize the cost $C(\vec{\theta})=\langle \psi(\vec{\theta})|H_G|\psi(\vec{\psi})\rangle$. Refs.~\cite{Xiaosi,bravo-prieto2019,huang2019near}  considered the Hamiltonian $H_G = A(\id - \dya{b})A\ad$ (which was also considered outside of the variational setting~\cite{subasi}).  The aforementioned cost can have gradients that vanish exponentially in the number of qubits $n$ (that is, a so-called barren plateau in the cost landscape). This problem can be mitigated by considering a local Hamiltonian with the same ground state~\cite{bravo-prieto2019} or by using a hybrid ansatz strategy~\cite{huang2019near}  where  $\ket{\psi(\vec{\theta})} = \sum_i \alpha_i \ket{\psi_i(\vec{\theta}_1)}$ with $\alpha_i$ being variational parameters. A study~\cite{bravo-prieto2019} was conducted with $n=10, \ldots, 30$ qubits for Ising-inspired linear systems and with $n=2,\ldots, 7$ qubit random (sparse) linear systems. The study showed that the time to solution scales logarithmically in $N$ (and also efficiently in condition number and solution precision) for these problems. Provided that larger systems display similar behavior, the observed heuristic scaling suggests that VQAs could potentially give an exponential speedup, analogous to HHL, for the QLSP.

\subsubsection{ Matrix-vector multiplication} Another related  problem is matrix-vector multiplication, that is to prepare a normalized state $\ket{x}$ such that $\ket{x}\propto A\ket{b}$ with normalized vector $\ket{b}$. When \mbox{$A=1-iH\delta t$}, then the problem becomes the task of Hamiltonian simulation. Similar to solving the QLSP, one constructs the Hamiltonian $H_{M}=\id-{A\left|b\right\rangle\left\langle b\right| A^{\dagger}}/{\| A\left|b\right\rangle \|^{2}}$, whose ground state is  $\ket{x}$ with zero energy~\cite{Xiaosi}. Here $\|A\ket{b}\|=\sqrt{\bra{b}A^\dag A\ket{b}}$ is the Euclidean norm. Given an approximate solution $\ket{\psi(\vec \theta^*)}$, one can lower bound the fidelity to the exact solution as $|\bra{\psi(\vec \theta^*)}x\rangle|^2\ge 1- \bra{\psi(\vec \theta^*)}H_{M}\ket{\psi(\vec \theta^*)}$, thus verify the solution's correctness whenever the cost function is small.

\subsubsection{ Non-linear equations} Non-linear equations are important to various fields, especially in the form of non-linear partial differential equations. However, mapping such equations onto quantum computers requires careful thought since the underlying mathematics of quantum mechanics is linear. To address this, a VQA for such non-linear problems was proposed in Ref.~\cite{lubasch2020variational}. The approach was illustrated for the time-independent non-linear Schr\"odinger equation, where the cost function is the total energy (sum of potential, kinetic, and interaction energies), and where the space was discretized into a finite grid. By using multiple copies of variational quantum states in the cost-evaluation circuit, this VQA can compute non-linear functions. 

An alternative approach has been proposed for non-linear differential equations that is based on using a set of basis functions rather than a finite grid~\cite{kyriienko2020solving}. First, the basis functions are encoded as non-linear feature maps (state preparation unitaries that are a function of the variables from the system). Next, a parameterized ansatz prepares a state that represents a linear combination of these basis functions. The corresponding function value is then output as an expectation value of an operator. Additionally, derivatives of this function are computed with the parameter shift rule. This method then optimizes a cost function that is minimized then the non-linear differential equation of interest is satisfied at a chosen set of points.

\subsubsection{ Factoring} Large-scale implementations of Shor's algorithm are not possible in the near term. Hence, a VQA for factoring as a potential near-term alternative was introduced in Ref.~\cite{anschuetz2019variational}. This proposal relies on the fact that factoring can be formulated as an optimization problem, and in particular, as a ground state problem for a classical Ising model. The authors used the QAOA to variationally search for the ground state. Their numerical heuristics suggest that a linear number of layers in the ansatz ($p\in \OC(n)$) leads to a large overlap with the ground state. 

\subsubsection{ Principal Component Analysis} An important primitive in data science is reducing the dimensionality of data with Principal Component Analysis (PCA). This involves diagonalizing the covariance matrix for a data set and selecting the eigenvectors with the largest eigenvalues as the key features of the data. Because the covariance matrix is positive semi-definite, one can store it in a density matrix, that is, in a quantum state, and then any diagonalization method for quantum states can be used for PCA. This idea was  exploited in Ref.~\cite{Lloyd2014} to propose a quantum algorithm for PCA. However,  quantum phase estimation and density matrix exponentiation were subroutines in this algorithm, making it non-implementable in the NISQ era. To potentially make this application more near-term, Ref.~\cite{larose2019variational} proposed a variational quantum state diagonalization algorithm, where the cost function $C(\vec{\theta})$ quantifies the Hilbert-Schmidt distance between the state $\tilde{\rho}(\vec{\theta})=U(\vec{\theta})\rho U(\vec{\theta})\ad$ and $\mathcal{Z}(\tilde{\rho}(\vec{\theta}))$, and where $\mathcal{Z}$ is the dephasing channel. This VQA outputs estimates of all the eigenvalues and eigenvectors of $\rho$, but it comes at the cost of requiring $2n$ qubits for an $n$ qubit state. This qubit requirement can be reduced with the VQA of Ref.~\cite{cerezo2020variational}, which requires only $n$ qubits. Here one exploits the connection between diagonalization and majorization to define a cost function of the form $C(\vec{\theta}) = \Tr[\tilde{\rho}(\vec{\theta}) H]$ where $H$ is a non-degenerate Hamiltonian. Due to Schur concavity, this cost function is minimized when $\tilde{\rho}(\vec{\theta})$ is diagonalized.

\subsection{ Compilation and unsampling}\label{sec:compilation}

A natural task that NISQ devices can potentially accelerate is the compiling of quantum programs. In quantum compiling, the goal is to transform a given unitary $V$ into native gate sequence $U(\vec{\theta})$ with an optimally short circuit depth. Quantum compiling plays a major role in error mitigation, as errors increase with circuit depth. Quantum compiling is a challenging problem for classical computers to perform optimally, due to the exponential complexity of classically simulating quantum dynamics. Hence, several VQAs have been introduced  that can potentially be used to accelerate this task~\cite{QAQC,heya2018variational,jones2018quantum,sharma2019noise,carolan2020variational}. These algorithms can be categorized as either Full Unitary Matrix Compiling (FUMC) or Fixed Input State Compiling (FISC), which respectively aim to compile the target unitary $V$ over all input states or for a particular input state. In Ref.~\cite{QAQC} a VQA for FUMC was presented, which uses  cost functions closely related to entanglement fidelities to quantify the distance between $V$ and $U(\vec{\theta})$. The proposal in Ref.~\cite{heya2018variational} also treats the FUMC case, but with an alternative approach to quantifying the cost using the average gate fidelity, averaged over many input and output states. The FISC case was treated in Ref.~\cite{jones2018quantum}, where the problem was reformulated as a ground state energy task, hence making the connection with VQE. The connection with VQE was also generalized to FUMC~\cite{sharma2019noise}, showing that variational quantum compiling, in general, is a special kind of VQE problem. Ref.~\cite{carolan2020variational} introduced and experimentally implemented a compiling scheme which can be thought of as FISC, although the architecture here is  focused on the application of unpreparing a quantum state. Finally, it is worth noting that  both FUMC and FISC exhibit resilience to hardware noise, in that the global minimum of the cost landscape is unaffected by various types of noise~\cite{sharma2019noise}. This noise resilience feature is crucial for the utility of variational quantum compiling for error mitigation, and we discuss this in more detail later.

\subsection{ Error correction}
Quantum Error Correction (QEC) protects qubits from hardware noise. Due to the large qubit requirements of QEC schemes, their implementation is beyond NISQ device capabilities. 
Nevertheless, QEC could still benefit NISQ hardware by suppressing the error to a certain extent and by combining it with other error mitigation methods. Specifically, conventional universal approaches for implementing QEC codes generally involve an unnecessarily long circuit that does not take into account the hardware structure or the type of noise. Hence, two VQAs have been introduced to solve these problems to automatically discover or compile a small quantum error-correcting code for any quantum hardware and any noise.

The Variational Quantum Error Corrector (QVECTOR) was first proposed to discover a device-tailored quantum error-correcting code for a quantum memory~\cite{johnson2017qvector}. For any $k$-qubit input state $\ket{\psi}=U_S\ket{\vec{0}}$, prepared by a unitary $U_S$ acting on a reference  state $\ket{\vec{0}}$, QVECTOR considered two parametrized circuits $V(\vec\theta_1)$ (on $n\ge k$ qubits) and $W(\vec\theta_2)$ (on $n+r$ qubits), which respectively encode the input logical state into $n$ qubits with $n-k$ ancillary qubits and realize recovery operations with $r$ ancillary qubits.  By sequentially applying encoding, recovery, and decoding on the input state, one obtains  an output $\rho_{out} = W(\vec\theta_1)V(\vec\theta_1)(\psi\otimes \ket{0}\bra{0}^{\otimes n-k+r}) V(\vec\theta_1)^\dag W(\vec\theta_1)^\dag$. Projecting the $n-k$ ancillary qubits back  to $\ket{0}\bra{0}$ and discarding the last $r$ ancillary qubits, one finds a quantum channel $\rho = \mathcal E(\vec\theta_1,\vec\theta_2)(\psi)$ on the input state $\psi$. The target of QVECTOR is to maximize the fidelity $\int_{\psi} d\psi F(\psi, \mathcal E(\vec\theta_1,\vec\theta_2)(\psi))$ between the output $\rho$ and the input $\psi$ averaged overall all $\psi$ or any $U_S$ that forms a unitary $2$-design. The solution will give the quantum circuit that maximally protects the input state. Numerical simulations showed that QVECTOR can find quantum codes that outperform existing ones~\cite{johnson2017qvector}.

Instead of discovering new device-tailored QEC codes, Ref.~\cite{xu2019variational} considered how to compile conventional QEC codes into a given quantum hardware with specific noise. Suppose one aims to implement the logical state $\ket{\psi}_L=\alpha\ket{0}_L+\beta\ket{1}_L$ with logical state basis $\{\ket{0}_L,\ket{1}_L\}$. Note that $\ket{\psi}_L$ is the ground state of the stabilizers $G_k$ as well as the logical operator $P=\ket{\psi}_L\bra{\psi}_L - \ket{\psi^{\perp}}_L\bra{\psi^\perp}_L$ with orthogonal state $\ket{\psi^{\perp}}_L$. Then one can construct a frustration-free Hamiltonian $H= - a_0P -\sum_{k\ge 1} a_k G_k$ with positive coefficients $a_0, a_k$, and with $\ket{\psi}_L$  the ground state with energy $E_G=-(a_0+\sum_{k\ge 1} a_k)$. One then uses a VQA to discover the circuit that implements $\ket{\psi}_L$ with a given hardware structure. Since the eigenstate energies are  know, the fidelity, $F$, of the discovered state can be bounded by $F\ge 1-(E-E_G)/a$ with the discovered energy $E$ and $a=\min\{a_0,a_k\}$. Numerical studies showed the encoding circuits for the five- and seven-qubit codes with different noisy hardware~\cite{xu2019variational}.

\subsection{ Machine learning and data science}\label{sec:ML}

Quantum machine learning (QML) generally refers to the tasks of using a quantum computer to learn patterns in quantum data with the goal of making accurate predictions on unknown, and unseen data~\cite{biamonte2017quantum}. Although an in-depth overview of QML is beyond the scope of this Review, we present several QML applications for which the VQA framework can be readily implemented. Specifically, here one learns a parametrized quantum circuit to solve a given task~\cite{farhi2018classification,mitarai2018quantum}. This connection between VQAs and (typical) QML applications shows that the lessons learned in one field can be of great use in the other, hence providing a close connection between these two fields. 

\subsubsection{ Classifiers} The classification of data is a ubiquitous task in machine learning.  Given training data of the form $\{ x^{(i)}, y^{(i)}\}$, where $x^{(i)}$ are inputs, and $y^{(i)}$ labels, the goal is to train a classifier to accurately predict the label of each input. Since a key aspect for the success of classical neural networks is their non-linearity, one can expect this property to also arise in a quantum classifier. As shown in Ref.~\cite{schuld2020circuit}, parametrized quantum circuits can support linear transformations and non-linearity can be exploited from the tensor product structure of a quantum system. More precisely, defining an input data dependent unitary $V(x)$, then the tensor product $V(x) \otimes V'(x)$ or 
the multiplication $V(x) V'(x)$ results in a non-linear function of the input data $x$. In this sense, the unitary $V(x)$ can be used as a quantum non-linear feature map,
where the Hilbert space can be exploited for a feature space~\cite{schuld2019quantum,havlivcek2019supervised}. Interestingly, the tensor network structure of quantum mechanics has even inspired  classical machine learning methods~\cite{stoudenmire2016supervised}.

Here, after embedding the input data $x$ into the quantum state, a linear transformation is performed using a parametrized quantum circuit, $U(\vec{\theta}) V(x) |\psi _0 \rangle$.
The cost function is then defined as the error between the true label and the expectation value of an 
easily measurable observable $A$, that is, $C(\vec{\theta}) = \sum_i [ y^{(i)} - \langle \psi _0 | V^{\dag}(x^{(i)}) U^{\dag}(\vec{\theta}) A U(\vec{\theta}) V(x^{(i)}) |\psi _0 \rangle ]^2 $.
This approach has been used in generalization and in classification tasks~\cite{mitarai2018quantum,schuld2020circuit}, with Refs.~\cite{mitarai2018quantum,lloyd2020quantum,schuld2020effect,perez2020data} discussing different ways of embedding classical data into quantum states (such as data re-uploading), and with Ref.~\cite{havlivcek2019supervised} showing an experimental demonstration of variational classification. 

Moreover, as shown in Refs.~\cite{havlivcek2019supervised,schuld2019quantum}, instead of using a parameterized unitary $U(\theta)$ one can use products of quantum feature vectors $\langle \psi_0 | V^{\dag}(x') V(x) |\psi _0 \rangle $ to perform a kernel method. Finally, the quantum kernel trick, which means that the dimensions of the quantum-enhanced feature space are larger than the number of data sets, has been demonstrated experimentally by using an ensemble nuclear spins~\cite{kusumoto2019experimental}.

\subsubsection{ Autoencoders} The autoencoder for data compression is an important primitive in machine learning. The idea is to force information through a bottleneck while still maintaining the recoverability of the data. As a quantum analog, Ref.~\cite{Romero} introduced a VQA for quantum autoencoding, with the goal of compressing quantum data. (see Refs.~\cite{wan2017quantum,verdon2018universal} for alternative approaches to quantum autoencoders.) The input to the algorithm is an ensemble of pure quantum states $\{p_{\mu},\ket{\psi_{\mu}}\}$ on a bipartite system $AB$ (here $p_\mu$ are real and positive coefficients such that $\sum_\mu,p_\mu=1)$. The goal is then to train an ansatz $U(\vec{\theta})$ to compress this ensemble into the $A$ subsystem, such that one can recover each state $\ket{\psi_{\mu}}$ with high fidelity from subsystem A. The $B$ subsystem is discarded and hence can be thought of as the `trash'. Given the close connection between data compression and decoupling~\cite{Romero}, the cost function is based on the overlap between the output state on $B$ and a fixed pure state. Recently, a local version of this cost function was also proposed and was shown to train well for large-scale problems~\cite{cerezo2020cost}. Moreover, in Ref.~\cite{cao2020noise}, the autoencoder scheme was generalized to mixed state and a noise-assisted algorithm was provided to improve the recovering fidelity for mixed/pure states. Quantum autoencoders have  seen experimental implementation on quantum hardware~\cite{pepper2019experimental}, and will likely be an important primitive in QML.

\subsubsection{  Generative models} The idea of training a parameterized quantum circuit for a QML implementation can also be applied for a generative model~\cite{verdon2017quantum,benedetti2019generative,du2020expressive}, which is an unsupervised statistical learning task with the goal of learning a probability distribution that generates a given data set. Let $\{x^{(i)}\}_{i=1}^{D}$ be a data set of size $D$ sampled from 
a probability distribution $q(x)$. Here one learns $q(x)$ as the parameterized probability 
distribution $p_{\theta}(x) =|\langle x | U(\vec{\theta}) |\psi _0 \rangle |^2 $ obtained by applying $U(\vec{\theta})$ to an input state and measuring in the computational basis, that is, it corresponds then to a quantum circuit Born machine~\cite{benedetti2019generative}. In principle one wishes to minimize the difference between the two distributions. However, since $q(x)$ is not available,  the cost function is defined by the negative log-likelihood 
$C(\vec{\theta})=- \frac{1}{D} \sum_{i} \log( p_{\theta}(x^{(i)}) )$.
In Ref.~\cite{benedetti2019generative} a variational framework for training quantum circuit Born machines was introduced and demonstrated for both classical data, such as the bars-and-stripes data set, and for synthetic data sets related to the preparation of cat states and coherent thermal states. In Ref.~\cite{liu2018differentiable}, an implicit generative model has been constructed by comparing the distance in the Gaussian kernel feature space. The representation power of the generative model has been investigated in Ref.~\cite{du2020expressive}.  Finally, it has been shown that quantum circuit Born machines can simulate the restricted Boltzmann machine and perform a sampling task that is hard for a classical computer~\cite{coyle2020born}.

\subsubsection{ Variational Quantum Generators} Generative Adversarial Networks (GANs) use two neural networks, a generator and discriminator, in competition. The generator aims to convince the discriminator that its output is coming from the true distribution associated with the training data. GANs play an important role in classical machine learning for applications such as image synthesis and molecular discovery. Ref.~\cite{romero2019variational} proposed a VQA for learning continuous distributions  which is meant to be a quantum version of GANs. Here one still considers classical data, but encoded into a quantum circuit. This encoding is followed by a variational quantum circuit that generates quantum states, which are then measured to produce a fake sample. This fake sample then enters either a classical discriminator or a quantum discriminator, and the cost function is optimized to minimize the discrimination probability with respect to real samples. The target application is to accelerate classical GANs using quantum computers.

\subsubsection{ Quantum Neural Network architectures} Several Quantum Neural Network (QNN) architectures have been proposed; for instance, Refs.~\cite{altaisky2001quantum,farhi2018classification,beer2020training} proposed perceptron-based QNNs. In these architectures each node in the neural network represents a qubit, and their connections are given by parameterized unitaries of the form in Eq.~\eqref{eq:general-ansatz-chain} acting on the input states. In addition, Ref.~\cite{cong2019quantum}  introduced Quantum Convolutional Neural Networks (QCNNs).  QCNNs have been used for error correction~\cite{cong2019quantum}, image recognition~\cite{franken2020explorations}, and to discriminate quantum state belonging to different topological phases~\cite{cong2019quantum}. Moreover, it has been shown that QCNNs and QNNs with tree tensor network architectures do not exhibit barren plateaus~\cite{pesah2020absence,zhang2020toward} (which will be discussed later), potentially making them a generically trainable architecture for large-scale implementations.

\subsection{ New frontiers}

In this section we discuss some exciting, recently proposed applications of VQAs. These applications highlight the fact that VQAs could be used to  understand and exploit the mathematical and physical structure of quantum states,  and  quantum theory in general. 

%In this section we discuss an exciting application of VQAs when dealing with quantum mechanical systems, that is, many VQAs could be used to  understand and exploit the mathematical and physical structure of quantum states,  and  quantum theory in general. 

\subsubsection{ Quantum foundations} NISQ computers will likely play an important role in understanding the foundations of quantum mechanics. In a sense, these devices offer experimental platforms to test foundational ideas ranging from quantum gravity to quantum Darwinism~\cite{zurek2009quantum}. For example, the emergence of classicality in quantum systems will be soon be a computationally tractable field of study due to the increasing size of NISQ computers. Along these lines, Ref.~\cite{arrasmith2019variational} proposed the Variational Consistent Histories (VCH) algorithm. Consistent Histories is a formal approach to quantum mechanics that has proven to be useful in studying the quantum-to-classical transition and quantum cosmology. In this formalism, interference between different paths (histories) as quantified in the decoherence functional~\cite{griffiths2003consistent}. The exponential number of terms in this decoherence functional makes the formalism computationally expensive on classical devices. VCH provides a way to prepare a density matrix representation of the entire functional, allowing one to efficiently examine the consistency of a set of histories. The application of standard VQAs to foundational situations can also provide a framework for new insights. For example, Ref.~\cite{holmes2020barren} showed that a Full Unitary Matrix Compiling (FUMC) strategy (discussed above) cannot efficiently learn a scrambling unitary. This result provides insight into the black hole information paradox as one would need to have a representation of a black hole's scrambling unitary in order to unscramble information from emitted Hawking radiation~\cite{hayden2007black}.

\subsubsection{ Quantum information theory} Another field that will likely see renewed interest due to NISQ computers is quantum information theory~\cite{wilde2013quantum}. For example, in Ref.~\cite{Romero} it was remarked that the quantum autoencoder algorithm could potentially be used to learn encodings and achievable rates for quantum channel transmission. Another area of research is using NISQ computers to compute key quantities in quantum information theory, such as the von Neumann entropy or distinguishability measures such as the trace distance. Although it is know that these problems are hard for general quantum states~\cite{watrous}, Ref.~\cite{cerezo2019variational} introduced a VQA to estimate the quantum fidelity between an arbitrary state $\sigma$ and a low-rank state $\rho$. Moreover, in Ref.~\cite{verdon2019quantumVQT} a VQA was introduced to learn modular Hamiltonians, which provides an upper bound on the von Neumann entropy of a quantum state. Here one attempts to variationally decorrelate a quantum state by minimizing the relative entropy to a product distribution, and hence this method is suited for states that can be easily decorrelated.

\subsubsection{ Entanglement Spectroscopy} Characterizing entanglement is crucial for understanding condensed matter systems, and the entanglement spectrum has proven to be useful in studying topological order. Several VQAs have been introduced to extract the entanglement spectrum of a quantum state~\cite{QSVD,larose2019variational,cerezo2020variational}. Since the entanglement spectrum can be viewed as the principlal components of a reduced density matrix, algorithms for PCA can be used for this purpose, including the VQAs discussed before. In addition, one can also use the variational algorithm for quantum singular value decomposition  introduced  in Ref.~\cite{QSVD}. These algorithms could potentially characterize the entanglement (and for example, topological order) in a ground state that was prepared by VQE, and hence different VQAs can be used together in a complementary manner.

\subsubsection{ Quantum metrology} Quantum metrology is a field where one seeks the optimal setup for probing a parameter of interest (for example a magnetic field) with minimal shot noise. In the absence of noise during the probing process, the analytical solution for the optimal probe state can be derived. However, when general physical noises are present, an analytical solution is hard to find. Variational-state quantum metrology variationally searches for the optimal probe state~\cite{koczor2019variational,kaubruegger2019variational,ma2020adaptive,beckey2020variational}. For state preparation, variational quantum circuits are used in Refs.~\cite{koczor2019variational,ma2020adaptive,beckey2020variational} whereas optical tweezer arrays are considered in Ref.~\cite{kaubruegger2019variational}. More concretely, one prepares a probe state with variational parameters, probes the magnetic field with physical noises, measures quantum Fisher information (QFI) as a cost function, and updates the parameters to maximize it. Note that since QFI cannot be efficiently computed, an approximation of QFI can be heuristically found by optimizing the measurement basis, or by computing upper and lower bounds on the QFI~\cite{beckey2020variational}. 

\section{ Challenges and potential solutions}\label{sec:4}

Despite the tremendous developments in the field of VQAs, there are still many challenges that need to be addressed to maintain the hope of achieving quantum speedups when scaling up these near-term architectures. Understanding the limitations of VQAs is crucial to developing strategies that can be used to construct better algorithms, prove certain guarantees on their performance, and even to build better quantum hardware.

\subsection{ Trainability}\label{sec:trainability}

\subsubsection{ Barren plateaus}\label{sec:BP}

\begin{figure}[t]
\centering
\includegraphics[width=1\columnwidth]{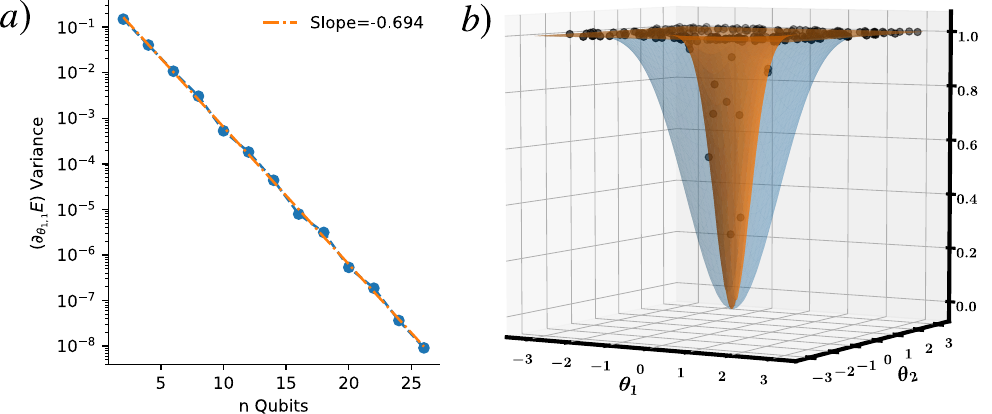}
    \caption{\textbf{Barren plateau (BP) phenomenon.} a. Variance of  the cost function partial derivative, $\text{Var}(\partial \theta_{1,1}E)$, for a particular parameter $\theta_{1,1}$ in the ansatz versus number of qubits ($n$). Results were obtained from a Variational Quantum Eigensolver implementation with a deep unstructured ansatz. The $y$-axis is on a log scale. As the number of qubits increases the variance vanish exponentially with the system size.  b. Visualization of the landscape of a global cost function which exhibits a BP for the quantum compilation implementation, . The orange (blue) landscape was obtained  for $n=24$ ($n=4$) qubits. As the number of qubits increases, the landscape becomes flatter. Moreover,  this cost also exhibits the narrow gorge phenomenon~\cite{cerezo2020cost}, where the volume of parameters leading to small cost values shrinks exponentially with $n$. Panel a is adapted from Ref.~\cite{mcclean2018barren}, CC BY 4.0;   Panel b is adapted from Ref. ~\cite{cerezo2020cost}, CC BY 4.0.  }
\label{fig:BP}
\end{figure}    

The so-called barren plateau (BP) phenomenon in the cost function landscape has received considerable attention as one of the main bottlenecks for VQAs. When a given cost function $C(\vec{\theta})$ exhibits a BP, the magnitude of its partial derivatives will be, on average, exponentially vanishing with the system size~\cite{mcclean2018barren}. As shown in Fig.~\ref{fig:BP} this has the effect of the landscape being essentially flat. Hence, in a BP  one needs an exponentially large precision to  resolve against finite sampling noise and determine a cost-minimizing direction, with this being valid independently of using a gradient-based~\cite{cerezo2020impact} or gradient-free optimization method~\cite{arrasmith2020effect}.  The exponential scaling in the precision due to BPs could erase a potential quantum advantage with a VQA, as its complexity would be comparable to the exponential scaling typically associated with classical algorithms. Hence, analyzing the existence of BPs in a given VQA is fundamental to preserve the hope of using it to achieve a quantum advantage.

The phenomenon of BPs was originally discovered in Ref.~\cite{mcclean2018barren} where it was shown that deep unstructured parametrized quantum circuits exhibit BPs when randomly initialized. The proof of this result relies on the fact that these unstructured ansatzes become $2$-designs when their depth grows polynomially with the number of qubits $n$~\cite{harrow2009random,brandao2016local}. One can view this phenomenon as stemming from the fact that the ansatz is problem-agnostic and hence needs to explore an exponentially large space to find the solution. Therefore, the probability of finding the solution when randomly initializing the ansatz is exponentially small.

The analysis of BPs was extended to shallow random layered ansatze~\cite{cerezo2020cost} where it was shown that the BP phenomenon is cost-function dependent: Global cost functions (when one compares operators or states living in exponentially large Hilbert spaces) exhibit BPs, whereas local cost functions (when one compares objects at the single-qubit level), exhibit gradients which vanish polynomially in $n$ so long as the circuit depth is at most logarithmic in $n$. This implies a connection between the locality and trainability and informs our intuition as to what types of cost functions might have to be avoided. These results have been numerically verified in Ref.~\cite{uvarov2020variational} and further extended in Ref.~\cite{uvarov2020barren}.  Here one can understand the presence of BPs for global costs as spanning from the fact that the Hilbert space grows exponentially with $n$ and hence the probability of two objects being close is exponentially small. 

In addition, it has been shown that BPs can arise in more general problems such as learning a scrambler~\cite{holmes2020barren} where for any choice of variational ansatz will lead to, with high probability, a BP. This again is due to the intrinsic randomness in a scrambler. In addition, the BP phenomenon has been studied when training randomly initialized  perceptron-based quantum neural  networks~\cite{sharma2020trainability,marrero2020entanglement}. Here, BPs arise from the significant amount of entanglement created by the perceptrons connecting large number of qubits in visible and hidden layers. Specifically, when tracing out the qubits in the hidden layers, the state of the visible qubits becomes exponentially close to being maximally mixed (due to concentration of measure), which makes it difficult to extract information from such a state. 

Although previous results rely on the randomness of the ansatz, there is a conceptually different phenomenon that can lead to BPs. Recently, it was shown in Ref.~\cite{wang2020noise} that noise can induce barren plateaus, regardless of the ansatz employed. Here, the presence of noise acting throughout the circuit progressively corrupts the state towards the fixed point of the noise model, usually the maximally mixed state~\cite{franca2020limitations}.  Such a phenomenon was shown to arise when the circuit depth needs to be linear (or larger) with the system size, meaning that it will affect many widely-used ansatzes.

\subsubsection{ Ansatz and initialization strategies}

Hand-in-hand with the theoretical progress on the analysis on the BP phenomenon, great effort has been dedicated to avoiding or mitigating the effect of BPs. The main strategy here has been to break the assumptions leading to BPs. In what follows we present two main approaches: parameters initialization and  choice of ansatz. 

\begin{itemize}
\item{Parameter initialization.} Randomly initializing an ansatz can lead to the algorithm starting far from the solution, near a local minima, or even in a region with barren plateaus. Hence, optimally choosing the seed for $\vec{\theta}$ at the beginning of the optimization is an important task. The importance of parameter initialization was made clear in Ref.~\cite{qaoaPerformanceLukin2018} where it was noted that the optimal parameters in QAOA exhibit persistent patterns. Based on these observations initialization strategies were proposed and which were heuristically shown to outperform randomly initialized optimizations. Additionally, in Refs.~\cite{grant2019initialization,wiersema2020exploring} an initialization strategy was developed specifically to address BPs in deep circuits. Here, one selects at random a subset of parameters in $\vec{\theta}$, and chooses the value of the remaining ones so that the circuit is a sequence of shallow unitaries that evaluates to the identity. The main idea behind this method is to reduce the randomness and depth of the circuit to break the assumption that the circuit approximates a $2$-design, a condition necessary for BPs to arise in deep ansatzes. Similar to the previous method, other schemes have been introduced to prevent BP by restricting the randomization of the ansatz. For instance, the proposal in Ref.~\cite{volkoff2020large} showed that correlating the parameters in the ansatz effectively reduces the dimension of the hyperparameter space and  can lead to large cost function gradients. In addition, Ref.~\cite{skolik2020layerwise} introduced a method where one uses layer-by-layer training: one initially trains shallow circuits and progressively adds components to the circuit. Whereas the latter guarantees that the number of parameters and randomness remains small for the first steps of the training, it has been shown~\cite{campos2020abrupt} that this method can lead to an abrupt transition in the ability of quantum circuits to be trained. Finally, a method was introduced in Ref.~\cite{verdon2019learning} where one pre-trains the parameters in the quantum circuits by using classical neural networks.

\item{Ansatz strategies.}  Another strategy for preventing BPs is using structured ansatzes which are problem-inspired. The goal here is to restrict the space explored by the ansatz during the optimization.   As discussed in the section on ansatzes, the UCC ansatz for VQE of the quantum alternating operator ansatz~\cite{qaoa2014,nasaQAOA2019} for optimization are problem-inspired ansatzes which are usually trainable even when randomly initialized. Other ansatz strategies include the proposals in Ref.~\cite{verdon2019learning} to learning a mixed state, where one leverages knowledge of the target Hamiltonian to create a Hamiltonian variational ansatz. In addition, Refs.~\cite{bharti2020iterative,bharti2020quantum} presented an approach where the ansatz for the solution is $\ket{\psi\left(\{c_\mu\}\right)} = \sum_\mu c_\mu \ket{\psi_\mu}$, for a fixed set of states $\{\ket{\psi_\mu}\}$ determined by the problem at hand. Here the optimization over the coefficients $\{c_\mu\}$ can be solved using a  quadratically constrained quadratic program. 
\end{itemize}
Finally, we remark that along with ansatz strategies there are other ways of potentially addressing BPs. These include optimizers tailored to mitigate the effect of BPs~\cite{anand2020natural}, local cost functions~\cite{cerezo2020variational}, or architectures such as the QCNN, which has been shown to avoid BPs~\cite{pesah2020absence}.

\subsection{ Efficiency}\label{sec:Efficiency}

Another requirement that must be met for VQAs to provide a quantum advantage is having an efficient way to estimate expectation values (and more general cost functions). The existence of BPs can exponentially increase the precision requirements needed for the optimization portion of VQAs, as discussed in the section on BPs, but even in the absence of such BPs  these expectation value estimations are not guaranteed to be efficient. Indeed, early estimations of resource requirements suggested that the number of measurements that would be required for interesting quantum chemistry VQE problems would be astronomical, hence addressing this issue is essential for realizing quantum advantage~\cite{wecker2015progress}. More reasonable resource estimates can be reached for restricted problems such as the  Hubbard model~\cite{cai2020hubbard,cade2020fermihubbard}. Although in principle one could always take projective measurements onto the eigenbasis of the operator in question, in general both the computational complexity of finding the required unitary, and the depth required to implement that transformation, may be intractable. However, given that arbitrary Pauli operators  are diagonalizable with one layer of single qubit rotations, it is common for the operators of interest (such as quantum chemistry Hamiltonians) to be expressed by their decomposition into such Pauli operators. That is,  $H=\sum_i c_i \sigma_i$, where $\{c_i\}$ are real coefficients and $\{\sigma_i\}$ are Pauli operators. The drawback of this approach is that, for many interesting Hamiltonians this decomposition contains many terms. For example, for chemical Hamiltonians the number of distinct Pauli strings scales as $n^4$ where $n$ is the number of orbitals (and thus qubits) for large molecules. In what follows we discuss several methods whose goal is to obtain measurement frugality in estimating the cost function.

\subsubsection{ Commuting sets of operators} In the interest of reducing the number of measurements required to estimate an operator expectation value, a number of methods have been proposed for partitioning sets of Pauli strings into commuting (simultaneously measurable)  subsets. The choice of the subsets is also of course non-unique and has been mapped onto the combinatorial problems of graph coloring~\cite{Jena2019,Crawford2019}, finding the minimum clique cover~\cite{verteletskyi2020measurement,Izmaylov2019,zhao2020measurement,Yen2019}, or finding the maximal flow in network flow graphs~\cite{Gokhale2019-2}, which makes it possible to import the heuristics and formal results from those problems. 

Perhaps the simplest approach to such a partitioning is to look for subsets that are qubit-wise commuting (QWC), which is to say that the Pauli operators on each qubit commute. Indeed, this was the first method introduced~\cite{mcclean2016theory}. 
However, whereas the QWC methods help reduce the number of operators, they do not change the asymptotic scaling for quantum chemistry applications, motivating more general commutative groupings to be considered. To this end, it has been shown that by considering general commutations (and increasing the number of gates of the circuit quadratically with $n$) the scaling of the number of measurements can be reduced to $n^3$~\cite{Jena2019,zhao2020measurement,Crawford2019,Izmaylov2019,Yen2019,Gokhale2019-2}. 

For using VQE on fermionic systems, this scaling can actually be brought down to either quadratic or,  for simpler cases, even linear~\cite{huggins2019efficient} in $n$. This significant improvement is found by considering  factorizations of the two-electron integral tensors, rather than working at the operator level. The success of this approach suggests that using background information on the problem  may significantly improve the measurement efficiency of estimating an expectation value.

\subsubsection{ Optimized sampling} In addition to reducing the number of individual operators that need to be measured, measurement efficiency can also be improved by carefully allocating the number of shots among the Pauli operators. Since operators with smaller coefficients will tend to contribute less to the overall variance, assigning the same number of shots to each operator is usually inefficient. Instead, the optimal approach~\cite{rubin2018application} is to give each Pauli operator a number of shots proportional to $|c_i| \sqrt{\text{Var}(\sigma_i)}$, where $c_i$ is the coefficient of the $i^\text{th}$ Pauli operator $\sigma_i$ and $\text{Var}(\sigma_i)$ is the variance of $\langle\sigma_i\rangle$. During an optimization where low precision steps may be allowed early on, this allocation can instead be performed randomly with probabilities proportional to $|c_i| \sqrt{\text{Var}(\sigma_i)}$. Making the allocation randomly in this way allows for unbiased estimates with as little as one shot, potentially significantly increasing the efficiency of the optimization~\cite{arrasmith2020operator}. Optimizing the sampling of the metric tensor has also been explored, with the conclusion that these costs need not be dominant in metric-aware VQAs~\cite{vanStraaten2020metricaware}.

\subsubsection{ Classical shadows} Another promising approach to efficient measurements is the construction of classical shadows~\cite{huang2020predicting}, also know as shadow tomography. In this approach, an approximate classical representation of the state (the classical shadow) is constructed by summing over the collection of states that a sequence of different measurements projects onto. These measurements are taken in the basis of randomly chosen strings so that a partial tomography of the state is completed. Combining the measurements in this way, each shot contributes to the estimation of each Pauli operator expectation value, resulting in a number of measurements that scales logarithmically with $n$. As with direct measurement approaches discussed above, this approach can also be further optimized by tuning the probability distribution for the Pauli operators that define the measurements to match the properties of the operator and state~\cite{hadfield2020measurements}. 

\subsubsection{ Neural network tomography} A different approach using partial tomography is to train an approximate restricted Boltzmann machine (RBM) representation of the desired quantum state~\cite{torlai2019precise}. This RBM is fitted using measurements of the  Pauli operators that are needed to directly estimate a given operator's expectation value, and so does not inherently reduce the number of operators to measure. However, by computing the expectation value on an approximate RBM instead of directly from measurements the sampling variance for a given number of shots is substantially reduced at the cost of introducing a small, positive bias~\cite{torlai2019precise}. 

\subsection{ Accuracy}~\label{sec:Accuracy}

One of the main goals for VQAs is to enable a practical use for NISQ devices. For this goal, VQAs provide a strategy to deal with hardware noise as they can potentially minimize quantum circuit depth. Moreover, as discussed below, error mitigation methods can be combined with VQAs to further improve accuracy. However, one can still ask what the impact of hardware noise will be on the accuracy of a VQA.

\subsubsection{ Impact of hardware noise}
   
There are multiple aspects of the impact of hardware noise: it could potentially slow down the training process, it could bias the landscape so that the noisy global optimum no longer corresponds to the noise-free global optimum, and it could affect the final value of the optimal cost.

\begin{itemize}   
\item{Effect of noise on training.}     The question of whether noise can help with the training process was posed in Ref.~\cite{gentini2019noise}. In practice, it is typical to observe that noise slows down the training. For example, it was heuristically observed that the noise-free cost achieves lower values with noise-free training than with noisy training~\cite{kubler2019adaptive,arrasmith2020operator,fontana2020optimizing}. As discussed in the section on BPs, the intuition behind this slowing down is that the cost landscape is flattened, and hence gradient magnitudes are reduced, by the presence of incoherent noise~\cite{xue2019effects,marshall2020characterizing,wang2020noise}. Moreover, gradients decay exponentially with the algorithm's depth, meaning that the deeper the circuit, the more it will be affected. This can be further understood from the fact that cost functions are typically extremized by pure states, and since incoherent noise reduces state purity, one expects this noise to erode the extremal points of the landscape~\cite{franca2020limitations}. The presence of noise-induced BPs and their effect on the trainability is one of the leading challenges for VQAs, with potential solutions being developing better quantum hardware or shorter-depth algorithms. It is worth remarking  that the results discussed here do not account for the use of error mitigation techniques, and the scope to which these could help is still an open question.

\item{Effect of noise on cost evaluation.}  In Refs.~\cite{wang2020noise,franca2020limitations} it was also shown that in the presence of local Pauli noise, the cost landscape concentrated exponentially with the depth of the ansatz around the value of the cost associated with the maximally mixed state. Whereas the proof of this exponential concentration of the cost was for general VQAs, some previous works had also observed this effect for the special case of the QAOA~\cite{xue2019effects,marshall2020characterizing}. The exponential concentration of the cost is of course important beyond the issue of trainability. Even if one is able to train, the final cost value will be corrupted by noise. There are certain VQAs where this is not an important issue (for example, in QAOA where one can classically compute the cost after sampling). However, for VQE problems, this is important, since one is ultimately interested in an accurate estimation of the energy. This emphasizes the importance of understanding to what degree error mitigation methods  can correct for this issue.
\end{itemize}

\subsubsection{ Noise resilience} 

One reason for the interest in VQAs is their ability to naturally overcome certain types of noise in hardware, especially in near-term implementations. This noise resilience is a crucial, non-trivial feature of VQAs.

\begin{itemize}
 \item{Inherent resilience to coherent noise.}  By construction, VQAs are insensitive to the specific parameter values, ultimately only sampling physical observables from the resulting state.  More specifically, if the physical implementation of a unitary results in a coherent error within the parameter space, or $U(\vec{\theta})$ actually results in $U(\vec{\theta} + \vec{\delta})$, then under mild assumptions the optimizer can calibrate this block unitary on the fly to improve the physical state produced.  This effect was first conjectured theoretically~\cite{mcclean2016theory} and later seen experimentally in superconducting qubits~\cite{o2016scalable}, where errors after the variational procedure were reduced in some cases by over an order of magnitude.  Success in this endeavor depends upon the ability to optimize faster than the drift of calibration in the device, and sufficient variational flexibility in the ansatz, but may continue to be effective even into the early fault-tolerant regime where coherent errors can be especially insidious.
    
\item{Inherent resilience to incoherent noise.} 
    It is an interesting question as to what degree incoherent noise, such as decoherence, random gate errors, and measurement errors, will impact VQAs. For example, it was shown that optimization in the presence of some noise channels can automatically move the state into subspaces that are resilient to those channels as an energetic trade-off~\cite{mcclean2017hybrid}.  However, one could operate under the assumption of perfect training (which may still be possible for either weak noise or shallow ansatzes), and ask whether the global optimum in the cost landscape is robust to such noise. This was the approach of Ref.~\cite{sharma2019noise}, where it was shown that VQAs for quantum compiling (see section on compilation), exhibit a special type of noise resilience known as Optimal Parameter Resilience (OPR). OPR is the notion that  global minima of the noisy cost function correspond to  global minima of the noise-free cost function. In this sense, if an algorithm exhibits OPR, then minimizing the cost in the presence of noise will still obtain the correct optimal parameters, and hence the optimal parameters are resilient. Since quantum compilation is a special case of VQE  the question still remains open to whether other VQAs exhibit this type of noise resilience for certain noise models.     A different type of noise resilience was analyzed in Ref.~\cite{kim2017noise} in the context of the holographic quantum simulation of many-body systems. Specifically, it was shown that under certain conditions the expectation values of local observables measures on the prepares ground-state are perturbed by, at most, a function that does not depend on the size, but rather only on the noise parameter. 
\end{itemize}

\subsubsection{Error mitigation}\label{sec:error-mitigation}

\begin{figure}
\centering
\includegraphics[width=1\columnwidth]{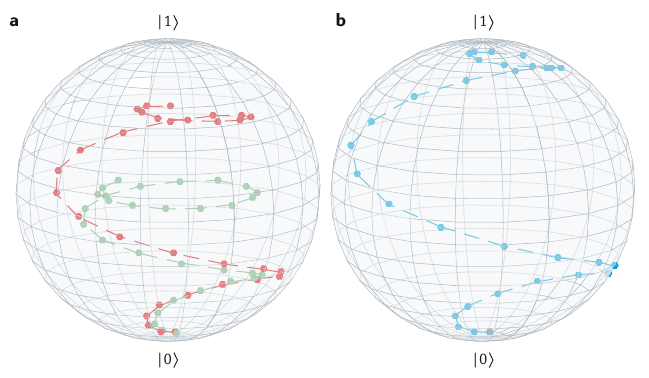}
    \caption{\textbf{Qubit trajectories on the Bloch sphere with the Zero-Noise Extrapolation (ZNE) technique.} The accuracy of a noisy quantum computer can be improved with the ZNE error mitigation method. a. Here, one repeats a given calculation with different levels of noise. The green  curve corresponds to a rotation on the Bloch sphere with a higher noise level than that leading to the red curve. b. Taking data from the red and green curves, ZNE can be used to estimate what the trajectory (blue) would be like in the absence of noise. Adapted from Ref.~\cite{kandala2019error}, Springer Nature Limited.}
\label{fig:ZNE}
\end{figure}

Quantum Error Mitigation (QEM) generally suppresses physical errors to expectation values of observables via classical post-processing of measurement outcomes~\cite{endo2020hybrid}. An intuitive, but powerful, example is the extrapolation method~\cite{li2017efficient,temme2017error}.  Even if the error rate cannot be reduced, in many cases it can be deliberately boosted, for example, as shown in Fig.~\ref{fig:ZNE}, by inserting additional noisy pulses or making gate operations longer, the quantum device undergoes more physical errors. Then, by obtaining  measurement outcomes at several noise levels and extrapolating them, one can estimate the error-free result using the so-called zero-noise extrapolation method. Due to the propagation of uncertainty, the variance of the error-mitigated result is amplified and hence one needs to have a larger sampling cost, which is the overhead of QEM. First, Richardson extrapolation was proposed~\cite{li2017efficient,temme2017error,kandala2019error}, and it was shown that single- and multi-exponential extrapolation work well for Markovian gate errors, with the latter subsequently shown to have very broad efficacy~\cite{endo2018practical,cai2020multiexponential}. In addition, extrapolation using least square fitting for several noise parameters has been proposed~\cite{otten2019recovering}. Furthermore, it has been observed that the extrapolation method can mitigate algorithmic errors that arise due to insufficiency in the number of time steps~\cite{endo2019mitigating}. 
  
Although extrapolation methods by design cannot fully mitigate physical errors~\cite{temme2017error,endo2018practical}, probabilistic error cancellation in theory can obtain unbiased expectation values by inverting the noise process with additional probabilistic gate operations (if a complete characterization of noise is provided). Note that since an inverse map of physical errors is generally unphysical, it is necessary to post-process measurement outcomes according to applied recovery operations.  In Ref.~\cite{temme2017error} this method was first introduced and in Ref.~\cite{endo2018practical} it was found that gate set tomography is a suitable noise characterization strategy, and a set of operations was proposed which can compensate for general Markovian errors. Furthermore, based on probabilistic error cancellation, stochastic error mitigation which works for general continuous systems such as analog quantum simulators and digital quantum computers was introduced~\cite{sun2020practical}.
        
A different approach to QEM relies on the classical simulability of near-Clifford circuits. The basic idea behind this approach is to compare the classically computed exact expectation values for near-Clifford circuits with their noisy counterparts evaluated on actual hardware~\cite{strikis2020learning,czarnik2020error,lowe2020unified}. Taking this approach can allow one to implement a probabilistic error mitigation protocol without needing to construct a full error model for an experiment~\cite{strikis2020learning}. Alternatively, one can perform a simple regression with this Clifford data to estimate how the observables have been affected and invert this regression to estimate desired noise-free expectation values~\cite{czarnik2020error}. Finally, zero-noise extrapolation can be merged with this regression to have an extrapolation to zero-noise whose form is tuned via the Clifford data, reducing the risk of blind extrapolations~\cite{lowe2020unified}.

Several additional QEM methods have been proposed. Symmetry verification is especially useful for ansatze that preserve symmetries such as particle and spin number~\cite{mcardle2019error,bonet2018low,otten2019noise}. Since physical errors break the symmetry, by measuring and ignoring the undesired case (similarly to error detection), one can mitigate physical noise. Unlike other QEM methods, symmetry verification can recover the quantum state itself. One can also take a post-processing approach using the information of the symmetry with a larger sample number~\cite{bonet2018low}. The use of symmetry verification to augment error extrapolation and probabilistic error cancellation was taken still further in Ref.~\cite{cai2020multiexponential}. 

In an alternative and complementary approach, the subspace expansion method was also shown to be useful for QEM in Ref.~\cite{mcclean2020decoding}. Here, using subspace expansion one can mitigate physical noise for eigenstates of the Hamiltonian as well as evaluating excited states because the state is expanded in a larger subspace. Note that this method works better for coherent noise than for stochastic noise. A distinct approach was introduced in Ref.~\cite{koczor2020exponential,huggins2020virtual} which comes at the cost of increasing the number of qubits. Here, by entangling and measuring $M$ copies of a noisy state $\rho$, one can compute expectation values with respect to the state $\frac{\rho^M}{\Tr[\rho^M]}$. Under the assumption that the principal eigenvector of $\rho$ is the desired state, this method can exponentially suppress errors with $M$. Finally, we remark that Ref.~\cite{bravyi2020mitigating} introduced a method to mitigate expectation values against correlated measurement errors, whereas Ref.~\cite{su2020error} implemented an error mitigation technique to suppress the effects  of photon loss for a Gaussian Boson sampling device.

\section{Opportunities for near-term quantum advantage}\label{sec:5}

VQAs are largely regarded as the best candidate for providing quantum advantage for practical applications. That is, it is expected that a VQA can solve a problem more efficiently than any classical state-of-the-art method. As discussed in the main text, tremendous effort has been dedicated to this goal with the development of efficient ansatz strategies, quantum-aware optimization methods, new VQAs, and error mitigation techniques. Although many challenges still remain to be addressed, such as the need for larger and better quantum devices, one can nevertheless pose the question as to what specific applications will provide the first quantum advantage for a practical scenario. In this section we discuss some of the most exciting possibilities where quantum advantage could arise.

\subsection{Chemistry and material sciences}
The ability to simulate and understand the static and dynamical properties of molecules and strongly correlated electronic systems is a fundamental task in many areas of science. For instance, this task is relevant in biology to understand protein folding dynamics, and in  pharmaceutical sciences one could analyze drug-receptor interactions to improve drug discovery capabilities~\cite{cao2018potential,cao2019quantum,outeiral2020prospects}. Similarly, analyzing the electronic structure of complex correlated materials is very important for studying high-temperature superconductivity or to analyze transition metal materials near a Mott transition. 

\subsubsection{Molecular structure} 
In the past few decades there have been great developments in the classical treatment of the structure of molecular systems. These include  approximate methods such as Hartree-Fock or density functional theory, or methods closely connected to quantum information, like the density matrix renormalization group approach that utilizes matrix product states as an ansatz~\cite{white1992density,chan2011density}.  However, even for these sophisticated approaches, systems of interest such as the FeMo cofactor are beyond the reach of an accurate description due to the entanglement structure of the electrons and orbitals.  The relevant electronic space that one needs to treat correlations accurately in for these systems is relatively modest, and for that reason,  these may be good targets for near-term quantum computers to play a role. As discussed in the main text the Variational Quantum Eigensolver algorithm~\cite{VQE} (and associated architectures)  have shown promising advances towards the goal of performing molecular quantum chemistry on quantum computers~\cite{mcardle2020quantum}, with large scale implementation already being executed~\cite{arute2020hartree}.

\subsubsection{Molecular dynamics} As for  the dynamics of chemical and other quantum systems, there have been a number of strides in evaluating or compressing these evolutions using variational approaches~\cite{li2017efficient,yuan2019theory}.  Much like variational principles connected to the ground state, there are a number of time-dependent variational principles that can be used to approximate time-dynamics. Here there are two timescales of interest.  The first is the electronic timescales over which electrons rearrange upon excitation.  The second, much slower than the first, is the rearrangement of nuclei that is induced by forces derived from the electrons in their respective configurations, excited or not.  Generally speaking, treating the detailed dynamics of the electrons accurately has been extremely challenging for classical approaches despite its relevance in phenomena related to photovoltaics and light-emitting diodes~\cite{bakulin2013charge,gross2000improving}.  The scale between the two  timescales has motivated the development of methods that treat them separately, often using a classical or semi-classical representation for the nuclei and quantum representation for the electrons~\cite{schmidt2008mixed}.  Variational methods can be applied incrementally in these cases, by stepping the electronic wavefunction forward with time-dependent variational principles~\cite{li2017efficient,yuan2019theory} and sampling the forces~\cite{o2019calculating} to move the nuclei classically, resulting in a Born-Oppenheimer type molecular dynamics.   Early test systems for quantum molecular dynamics often include photo-dissociation reactions and conical interactions of small molecular systems~\cite{tully1971trajectory}.  Ultimately, these methods may help unlock proton-coupled electron transfer mechanisms~\cite{weinberg2012proton} in proteins and help with the design of novel organic photovoltaics~\cite{bakulin2013charge} and related systems.

\subsubsection{Materials science} Classical methods for materials simulations usually use density-functional theory coupled with approximation methods, such as the  local density approximation~\cite{kohn1965self} to tackle weakly correlated materials. However, many effects arising from strongly correlated systems are beyond the reach of such classical methods. Since long-term algorithms for material simulation require phase estimation~\cite{bauer2016hybrid,babbush2018encoding,berry2018improved}, these lie beyond the scope of near-term devices. In contrast, near-term VQAs for analyzing strong correlation problem are aimed at reducing the circuit depth by using smart initializations~\cite{dallaire2018low}, or by optimizing the circuit structure itself~\cite{grimsley2019adaptive,tang2019qubit}.

\subsection{Nuclear and particle physics}

\subsubsection{Nuclear physics} Similar to the chemistry applications discussed above, VQAs have the potential to convey a quantum advantage in studying nuclear structure and dynamics. The most studied potential contribution is the utility of the VQE method to find nuclear ground states. This was first demonstrated for computing the deuteron ($^2$H) binding energy~\cite{dumitrescu2018cloud}, and has been extended to other light nuclei such as the triton ($^3$H), $^3$He, and an alpha particle ($^4$He)~\cite{lu2019simulations}. Additionally, using VQE to prepare the ground state of a triton has been an initial step as a demonstration of simulating neutrino-nucleon scattering~\cite{roggero2020quantum}. Considering these low-energy applications along with the general progress towards studying higher energy nuclear interactions (quantum chromodynamics) via VQA lattice gauge theory approaches (discussed below) shows that VQAs have the potential to provide a significant advantage over classical methods for nuclear physics.

\subsubsection{Particle physics} In particle physics many analytical tools have been developed to describe and study theories, but there are many areas that remain intractable. In particular, the study of important gauge theories such as quantum chromodynamics is often handled by mapping the problem onto a lattice to allow for numerical studies. One of the major drawbacks of such Lattice Gauge Theories (LGTs) for classical computation is that they exhibit the sign problem and as a result are usually not classically simulable. Although large scale, fault-tolerant quantum computers will eventually be able to handle this difficulty~\cite{bender2018digital,banuls2020simulating}, there is also the potential for achieving a significant quantum advantage in this area with VQAs in the Noisy Intermediate-Scale Quantum (NISQ) era~\cite{preskill2018simulating}. Advances in this direction include work on VQAs for LGT  simulation~\cite{kokail2019self} and variational determinations of mass gaps, Green's functions, and running coupling constants~\cite{endo2020calculation,mishra2020quantum,paulson2020towards}. In addition, an approach using a VQA to determine interpolation operators to accelerate classical LGT computations has been proposed~\cite{avkhadiev2020accelerating}. Finally, the impacts of decoherence by hardware noise on LGT calculations have been studied, finding that gauge violations caused by decoherence only grow linearly at short times, suggesting that short depth approaches may be possible~\cite{halimeh2020fate}. Taken together, these results show that studying LGTs is a viable candidate for NISQ quantum advantage. 

\subsection{Optimization and machine learning}

Although it is natural to consider that VQAs can bring an advantage on tasks which are inherently quantum in nature, the prospect of using quantum algorithms to solve classical problems is also an exciting one. Generally, one here aims to use the large dimension of the Hilbert space to encode big problems or large amounts of data, with the premise that the quantum nature of the algorithm (such as coherence or entanglement between qubits~\cite{sharma2020reformulation})  helps in speeding up a given task.

\subsubsection{Optimization} Many optimization problems can be encoded in relatively simple mathematical models such as the Max-Cut~\cite{qaoaMaxCut2018} or the Max-Sat~\cite{Lin2016sat} problems. These include tasks such as electronic circuitry layout design, state problems in statistical physics~\cite{barahona1988application}, and even automotive configuration~\cite{kuchlin2000proving}. Applying Quantum Approximate Optimization Algorithm (QAOA) to classical optimization problems is widely considered to be one of the leading candidates for achieving quantum advantage on NISQ devices~\cite{crooks2018performance}. There are several reasons for this optimism. QAOA has provable performance guarantees~\cite{qaoa2014, farhi2014qaoa_for_bounded} for $p=1$. In general, even $p=1$ QAOA ansatz cannot be efficiently simulated on any classical device~\cite{farhi2016supremacy}. At the same time, QAOA performance can only improve by increasing $p$. It was also shown that `bang-bang' evolution that motivates QAOA ansatz is the optimal approach given fixed quantum computation time~\cite{Yang2017optimizing}. However, there are problems for which a shallow QAOA ansatz does not perform well~\cite{hastings2019classical, bravyi2019obstacles} suggesting that $p$ may have to grow with the problem size. Larger $p$ requires improvements in the parametrization and optimization~\cite{qaoaPerformanceLukin2018}. Similarly to quantum chemistry, large scale experiments of QAOA have already been implemented~\cite{arute2020quantum}.

\subsubsection{Machine Learning} In the past few decades, the use of machine learning has become common in most, if not all, areas of science. Although the problem of loading classical data on quantum computers is still an active topic of research, there has been significant efforts put forward to use quantum algorithms for machine learning applications~\cite{schuld2015introduction,schuld2019quantum,romero2019variational,biamonte2017quantum}. For instance, it has been shown that quantum neural networks can achieve a significantly higher capacity, as measured by the effective dimension, than comparable classical neural networks~\cite{abbas2020power}, implying that the former can express a broader class of functions than the latter. Moreover, it has also been pointed out that quantum algorithms can outperform classical ones in deep  learning problems~\cite{wiebe2014quantum}, potentially provide exponentially better ability to generalize when trained to predict the outcome of physical processes~\cite{huang2021information}, and more recently a VQA has been proposed for deep reinforcement learning~\cite{chen2020variational}. An exciting prospect for using quantum neural networks is that certain architectures are immune to barren plateaus, and hence are trainable even for large problems~\cite{pesah2020absence,zhang2020toward}.   

\section{ Outlook}~\label{sec:7}

In the quest for quantum advantage, analytical and heuristic scaling analysis of VQAs will be increasingly important. Better methods to analyze VQA scalability are anticipated in the future. This will likely include both gradient scaling and other scaling aspects, such as the density of local minima and the shape of the cost landscape. These fundamental results will help to guide the search for quantum advantage.

At the same time, the future will also see an improved toolbox for VQAs. Quantum-aware optimizers will exploit knowledge gained about the cost landscape. These improved optimizers will mitigate the impacts of small gradients and avoid local minima to facilitate rapid training of the parameters in VQAs. Moreover, commercial software packages will streamline the testing of VQAs and further speed up the parameter optimization~\cite{bergholm2018pennylane,broughton2020tensorflow,luo2020yao}.

Application-specific ansatzes will continue to be developed. Better ansatzes will enhance gradient magnitudes to improve trainability and they may also reduce the impact of noise on VQAs. This will likely include adaptive ansatz strategies, which appear promising. Hybrid quantum-classical models~\cite{verdon2019quantumVQT} are a natural extension of VQAs where one parameterizes both a classical (for example, neural network) and quantum ansatz, and such models could also facilitate near-term applications.

New error mitigation strategies are anticipated in the future. These will be crucial for obtaining accurate results from VQAs and will improve accuracy by orders of magnitude. Error mitigation will be hard-coded into cloud-based quantum computing platforms, to allow uses to obtain accurate results with ease. 

The future will also see better quantum hardware become available, both in terms of qubit count and noise levels. VQAs will certainly benefit from such improved hardware. Moreover, VQAs will play a central role in benchmarking the capabilities of these new platforms.

In the near future, VQAs will likely see a shift from the proposal and development phase to the implementation phase. Researchers will aim to implement
larger, more realistic problems with VQAs instead of toy problems. These implementations will incorporate multiple state-of-the-art strategies for enhancing VQA performance. Combining strategies for improving the accuracy, trainability, and efficiency of VQAs will test their ultimate capabilities and will push the boundaries of NISQ devices, with the grand vision of obtaining quantum advantage.

In the more distant future, VQAs will even find use even when the fault-tolerant era arrives. Transitioning from estimating expectation values from Hamiltonian averaging to phase estimation may be an important component here~\cite{wang2019accelerated}. QAOA may be a good candidate VQA to find usage in the fault-tolerant era, albeit with caveats about the overhead~\cite{sanders2020compilation}. Strategies that address challenges in the NISQ era, such as keeping circuit depth shallow and avoiding barren plateaus, could still play a role in the fault-tolerant era. Therefore, current research on VQAs will likely remain useful even when fault-tolerant quantum devices arrive.

\bibliography{ref.bib}

\section*{Acknowledgements}

%MC
MC is thankful to Kunal Sharma for helpful discussions. MC was initially supported by the Laboratory Directed Research and Development (LDRD) program of
Los Alamos National Laboratory (LANL) under project number 20180628ECR, and later supported by the Center for Nonlinear Studies at LANL. 
%AA
AA was initially supported by the LDRD program of
LANL under project number 20200056DR, and later supported by the by the U.S. Department of Energy (DOE), Office of Science, Office of High Energy Physics QuantISED program under under Contract Nos.~DE-AC52-06NA25396 and KA2401032.
%SB
SCB acknowledges financial support from EPSRC Hub grants under the agreement numbers EP/M013243/1 and EP/T001062/1, and from EU H2020-FETFLAG-03-2018 under the grant agreement No 820495 (AQTION). 
%SE
SE was supported by MEXT Quantum
Leap Flagship Program (MEXT QLEAP) Grant Number
JPMXS0120319794, JPMXS0118068682 and JST ERATO Grant Number JPMJER1601.
%KF
KF was supported by JSPS KAKENHI Grant No. 16H02211,  JST ERATO JPMJER1601, and JST CREST JPMJCR1673.
%KM
KM was supported by JST PRESTO Grant No. JPMJPR2019 and JSPS KAKENHI Grant No. 20K22330.
KM and KF were also supported by MEXT Quantum Leap Flagship Program (MEXT QLEAP) Grant Number JPMXS0118067394 and JPMXS0120319794. 
%XY
XY acknowledges support from the Simons Foundation.
%LC
LC was initially supported by the LDRD program of LANL under project number 20190065DR, and later supported by the U.S. DOE, Office of Science, Office of Advanced Scientific Computing Research under the Quantum Computing Application Teams (QCAT) program. 
%PJC
PJC was initially supported by the LANL ASC Beyond Moore's Law project, and later supported by the U.S. DOE, Office of Science, Office of Advanced Scientific Computing Research, under the Accelerated Research in Quantum Computing (ARQC) program. Most recently, MC, LC, and PJC were supported by the Quantum Science Center (QSC), a National Quantum Information Science Research Center of the U.S. Department of Energy (DOE).

\section*{Author contributions}
All authors have read, discussed and contributed to the writing of the manuscript.

\section*{Competing interests}
The authors declare no competing interests.

\section*{Key points:} 
\begin{itemize}
\item Variational quantum algorithms (VQAs) are the leading proposal for achieving quantum advantage using near-term quantum computers.

\item VQAs have been developed for a wide range of applications including finding ground states of molecules, simulating dynamics of quantum systems, and solving linear systems of equations, among others.

\item VQAs share a common structure, where a task is encoded into a parameterized cost function that is evaluated using a quantum computer, and a classical optimizer trains the parameters in the VQA.

\item The adaptive nature of VQAs is well suited to handle the constraints of near-term quantum computers.

\item Trainability, accuracy, and efficiency are three challenges that arise when applying VQAs to large-scale applications, and strategies are currently being developed to address these challenges.
\end{itemize}

\end{document}